\documentclass[11pt]{article} 

\usepackage[ansinew]{inputenc}
\usepackage{amsmath, amssymb, graphics, amsthm}
\usepackage{epsfig}
\usepackage{color}
\usepackage{undertilde}
\usepackage{fancyhdr} 
\usepackage{bbold}
\usepackage{slashed}

\newcommand{\m}{\mbox{}}

\makeatletter
\@addtoreset{equation}{section}
\makeatother

\newcommand{\be}{\begin{equation}}
\newcommand{\ee}{\end{equation}}
\newcommand{\ba}{\begin{eqnarray}}
\newcommand{\ea}{\end{eqnarray}}

\title{{\sf Towards Loop Quantum Supergravity (LQSG)}\\
{\sf I. Rarita-Schwinger Sector}} 
\author{
{\sf N. Bodendorfer}$^1$\thanks{{\sf 
norbert.bodendorfer@gravity.fau.de}},
{\sf T. Thiemann}$^{1,2}$\thanks{{\sf 
thomas.thiemann@gravity.fau.de,
tthiemann@perimeterinstitute.ca}},
{\sf A. Thurn}$^1$\thanks{{\sf 
andreas.thurn@gravity.fau.de}}\\
\\
{\sf $^1$ Inst. for Theoretical Physics III, FAU Erlangen -- N\"urnberg,}\\
{\sf Staudtstr. 7, 91058 Erlangen, Germany}\\
\\
{\sf and}\\
\\
{\sf $^2$ Perimeter Institute for Theoretical Physics,}\\ 
{\sf 31 Caroline Street N, Waterloo, ON N2L 2Y5, Canada}
}
\date{{\small\sf \today}}

\begin{document} 

\maketitle

{\sf

\begin{abstract}

In our companion papers, we managed to derive a connection formulation of Lorentzian
General
Relativity in $D+1$ dimensions with compact gauge group SO$(D+1)$ such that the connection is Poisson commuting, which implies that Loop Quantum Gravity quantisation methods apply. 
We also provided the coupling to standard matter. 

In this paper, we extend our methods to derive a connection formulation of a large class of Lorentzian signature Supergravity theories, in particular $11d$ SUGRA and $4d$,  $N=8$ SUGRA,
which was in fact the motivation to consider higher dimensions. Starting from a Hamiltonian formulation in the time gauge which yields a Spin$(D)$ theory, 
a major challenge is to extend the internal gauge group to Spin$(D+1)$ in presence of the Rarita-Schwinger field.  This is non trivial because SUSY typically requires the Rarita-Schwinger
field to be a Majorana fermion for the Lorentzian Clifford algebra and Majorana representations of the Clifford algebra are not available in the same spacetime dimension for both Lorentzian and 
Euclidean signature.

We resolve the arising tension and provide a background independent representation of the non trivial 
Dirac antibracket $^\ast$-algebra for the Majorana field which significantly differs from 
the analogous construction for Dirac fields already available in the literature. 

\end{abstract}

}

\newpage

{\large \bf Notation}

\ba
D &~& \text{spatial dimension} \\
d = D+1 &~& \text{spacetime dimension} \\
s,\zeta &~& \text{spacetime, internal signature} \\
\eta_{IJ} &=& \text{diag}(\zeta, 1,1,...) \\
I,J,... &~&\text{SO$(D,1)$ or SO$(D+1)$ indices (short: SO$(\eta)$)} \\
i,j,... &~&\text{SO$(D)$ indices} \\
\alpha,\beta,... &~&\text{spinor indices} \\
\mu,\nu,... &~&\text{spacetime indices} \\
a,b,... &~&\text{spatial indices} \\
\left\{ \gamma^I, \gamma^J \right\} &=& 2 \eta^{IJ} \\
(\gamma^{I})^{ \dagger} &=& \eta^{II} \gamma^{I} ~~~~~~~~ \text{(no summation here)} \label{eq:GammaAdjoint} \\
\gamma^{IJ...K} &:=& \gamma^{[I} \gamma^{J} ... \gamma^{K]} ~\text{(with total weight one)} \\
\gamma^{\perp} &:=& n_{\mu} \gamma^{\mu}\\
\Sigma^{IJ} &:=& - \frac{i}{2} \gamma^{IJ} \\
\nabla_a(A)~ \chi &:=& \partial_a \chi + \frac{i}{2} A_{aIJ} \Sigma^{IJ} \chi \\
i \left[\Sigma^{IJ}, \Sigma^{KL}\right] &=& \eta^{LJ} \Sigma^{KI} - \eta^{LI} \Sigma^{KJ} - \eta^{KJ} \Sigma^{LI} + \eta^{KI} \Sigma^{LJ} \\
\bar{\chi} &:=& \chi^{\dagger} \gamma^0 \\
\bar{\chi} &=& \chi^T C ~~~~~~\text{Majorana Condition} \label{eq:MajoranaCondition}
\ea
Note that the choice (\ref{eq:GammaAdjoint}) is always possible \cite{NieuwenhuizenAnIntroductionTo}. Furthermore, note that in a Majorana representation the spinors will be real and then, from the Majorana condition (\ref{eq:MajoranaCondition}) follows $\gamma^0 = C$.

\newpage

\tableofcontents

\newpage
    
\section{Introduction}

During the years after the discovery of $D+1=3+1$ Supergravity by Freedman, Ferrara, and van Nieuwenhuizen in 1976 \cite{FreedmanProgressTowardsA}, there has been a lot of activity in the newly formed field of Supergravity, driven by the hope to construct a theory of quantum gravity without the shortcoming of perturbative non-renormalisability. Werner Nahm classified in 1977 all possible Supergravities, arriving at the result that, under certain assumptions, $d=11$ was the maximal number of Minkowski signature spacetime dimensions in which Supergravities could exist \cite{NahmSupersymmetriesAndTheir}. In the following year, $d=11$ Supergravity was constructed by Cremmer, Julia and Scherk \cite{CremmerSupergravityTheoryIn} in order to obtain $d=4$, $N=8$ maximal Supergravity by dimensional reduction. Various forms of Supergravity were derived in dimensions $d \leq 11$ and relations among them were discovered in the subsequent years  \cite{SalamSupergravitiesInDiverse}.

While the initial hope linked with perturbative Supergravity was vanishing due to results suggesting its non-renormalisability \cite{DeserNonrenormalizabilityOfLast} and the community turned to Superstring theory, a new candidate theory for quantum gravity, Loop Quantum Gravity (LQG), started to emerge after Ashtekar discovered his new variables in 1986 \cite{AshtekarNewVariablesFor}. During the next 10 years, the initially complex variables were cast into a real form by Barbero \cite{BarberoRealAshtekarVariables}, rigorous techniques for the  construction of a Hilbert space were developed \cite{AshtekarRepresentationsOfThe, AshtekarRepresentationTheoryOf, AshtekarDifferentialGeometryOn, AshtekarProjectiveTechniquesAnd, MarolfOnTheSupport, AshtekarQuantizationOfDiffeomorphism} and a representation of the constraint operators on the Hilbert space was defined \cite{ThiemannQSD1}. The strengths of the theory are, among others, its entirely non-perturbative and background-independent formulation as well as the suggested appearance of a quantum geometry at the Planck scale. It is therefore in a sense dual to the perturbative descriptions coming from conventional quantum (Super)gravities and Superstring- / M-theory and it would be very interesting to compare and merge the results coming from these two different approaches to quantum gravity. The main conceptual obstacle in comparing these two methods of quantisation has been the spacetime they are formulated in. While the Ashtekar-Barbero variables are only defined in $3+1$ dimensions, where also an extension to supersymmetry exists, Superstring- / M-theory favours $9+1$ / $10+1$ dimensions and is regarded as a quantisation of the respective Supergravities. It is therefore interesting to study quantum Supergravity as a means of probing the low-energy limit of Superstring- / M-theory with different quantisation techniques, both perturbative and non-perturbative. A somewhat different approach has been taken in \cite{ThiemannTheLQGString1}, where the closed bosonic string has been quantised using rigorous background-independent techniques, resulting in a new solution of the representation problem which differs from standard String theory. Also, the Hamiltonian formulation of the algebraic first order bosonic string and its relation to self-dual gravity have been recently investigated in \cite{FairbairnCanonicalAnalysisOf, FairbairnEquivalenceOfThe}.

Apart from contact with Superstring- / M-theory, new results from perturbative $d=4$, $N=8$ Supergravity \cite{BernUltraviolettBehaviorOf, GreenNonrenormalizationConditionsIn, KalloshOnUVFiniteness} suggest that the theory might be renormalisable, contrary to prior believes. It is therefore interesting in its own right to study the loop quantised $d=4$, $N=8$ theory and compare the results with the perturbative expansion.

The main obstacle in the connection formulation of General Relativity with or without SUSY is represented by the gravitational variables. Starting from the plain Einstein-Hilbert-term in a (Super)gravity action, one obtains the ADM variables $q_{ab}$ and $P^{ab}$ \cite{ArnowittTheDynamicsOf}. In order to incorporate fermions, it is mandatory to use tetrads and their higher dimensional analogues (vielbeins) to construct a representation of the curved spacetime Dirac (Clifford) algebra. At this point, it is convenient to use the time gauge \cite{SchwingerQuantizedGravitationalField} to obtain a densitised $D$-bein $E^a_i$ and its canonically conjugate momentum $K_a^i$ as canonical variables. Since the time gauge fixes the boost part of the internal gauge group SO$(1,D)$, we are left with a SO$(D)$ gauge theory. The problem however is that the Gau{\ss} constraint generating the internal rotations transforms  $E^a_i$ and $K_a^i$ as internal vectors, and thus not as a connection and a vector. It was shown in \cite{BTTI, BTTIV} that starting from this formulation, one can construct a canonical transformation which leads to a formulation in terms of a SO$(D+1)$ connection $A_{aIJ}$ and its canonically conjugate momentum $\pi^{aIJ}$, where it was necessary to enlarge the dimension of the internal space by one space-like dimension. 

The purpose of this paper is to generalise this transformation to Supergravity theories. The problem arising in these generalisations are not so much linked to the appearance of additional tensor fields and spin $1/2$ fermions, but to the Rarita-Schwinger field which obeys a Majorana condition. It is well known \cite{DeserHamiltonianFormulationOf} that in order to have simple and metric independent Poisson brackets for the Rarita-Schwinger field $\psi^\alpha_a$, one should use half-densitised internally projected fields $\phi^\alpha_i := \sqrt[4]{q} e^a_i \psi^\alpha_a$. This field redefinition has to be changed in order to work in the new internal space, more specifically we have to ensure that the number of degrees of freedom still matches by imposing suitable constraints. Also, the Majorana conditions are sensitive to the dimensionality and signature of spacetime. We thus have to ensure that no inconsistencies arise when using SO$(D+1)$ instead of SO$(1,D)$ as the internal gauge group. Concretely, this will be achieved in dimensions where Majorana representation for the $\gamma$-matrices exists, which covers many interesting Supergravity theories ($d=4, 8, 9, 10, 11$).

The presence of additional tensors, vectors, scalars and spin $1/2$ fermions in various 
SUGRA theories does not pose any problems for this classical canonical transformation. 
However, we must provide background independent representations for these fields
in the quantum theory which, to the best of our knowledge, has not been done yet
for all of them. As an example, in our companion 
paper \cite{BTTVII} we consider the quantisation of Abelian $p$-form fields such as the 3-index
photon present in $11d$ SUGRA with Chern-Simons term.  
Scalars, fermions and connections of compact, possibly non-Abelian, gauge groups have already been treated in \cite{ThiemannKinematicalHilbertSpaces}.\\
\\
The article is organised as follows:\\
\\ 
Section 2 is subdivided into two parts. In the first we review  prior work on canonical Supergravity theories in various dimensions and identify their common structural elements. We also mention the basic difficulties in our goal to 
match these canonical formulations to the reformulations \cite{BTTI,BTTII} of the graviton sector.
In the second we display canonical Supergravity explicitly in the time
gauge paying  special attention to the Rarita-Schwinger sector.  

Section 3 is also subdivided into two parts. In the first we display the symplectic structure
of the Rarita-Schwinger field in the time gauge in convenient variables which will be crucial for a later
quantisation of the theory. In the second, following the strategy in \cite{BTTI,BTTIV} we will 
perform an extension of the phase space subject to additional second class constraints 
ensuring that we are dealing with the same theory while the internal gauge group can 
be extended from SO$(D)$ to SO$(D+1)$.

In section 4 we construct a representation of the Dirac anti bracket geared to Majorana
spinor fields rather than Dirac spinor fields.

In section 5 we show that our formalism easily extends without additional complications 
to chiral Supergravities (Majorana-Weyl spinors) and to spin $1/2$ Majorana
fields which are present in some Supergravity theories.

In section 6 we summarise and conclude.    

Finally, in the appendix, we supply the details of the formulation of higher dimensional 
connection General Relativity with linear simplicity constraints in terms of 
a normal field $N^I$ which is convenient in order to resolve the afore mentioned tension 
between SO$(1,D)$ and \mbox{SO$(D+1)$} Majorana spinors and we provide a Hilbert space 
representation of the normal field sector.

\section{Review of Canonical Supergravity}

In the first part of this section we summarise the status of canonical Supergravity and 
its quantisation. In the second we display the details of the theory to the extent we need 
it which will settle the notation.

\subsection{Status of Canonical Supergravity}

Hamiltonian formulations of Supergravity are a tedious business due to the complexity of the Lagrangians and the appearance of constraints. Nevertheless, the canonical structure emerging is very similar for the explicitly known Hamiltonian formulations. To the best of our knowledge, the $D+1$ split for $D \geq 3$ has been explicitly performed for $D+1=3+1$, $N=1$ \cite{DeserHamiltonianFormulationOf, FradkinHamiltonianFormalismQuantization, PilatiTheCanonicalFormulation, SenjanovicHamiltonianFormulationAnd}, $D+1=9+1$, $N=1$ \cite{HenneauxHamiltonianFormulationOf}, and $D+1=10+1$, $N=1$ \cite{DiazHamiltonianFormulationOf}. The algebra of constraints of $D+1=3+1$ Supergravity was first computed by Henneaux \cite{TeitelboimSupergravityAndSquare} up to terms quadratic in the constraints \cite{HenneauxPoissonBracketsOf}. The same method was applied by Diaz \cite{DiazConstraintAlgebraIn} to $D+1=10+1$ Supergravity, also neglecting terms quadratic in the constraints. Sawaguchi performed an explicit calculation of the constraint algebra of $D+1=3+1$ Supergravity in \cite{SawaguchiCanonicalFormalismOf} where a term quadratic in the Gau{\ss} constraint appears in the Poisson bracket of two supersymmetry constraints. The constraint algebra for $D+1=9+1$, $N=1$ Supergravity coupled to supersymmetric Yang-Mills theory was calculated by de Azeredo Campos and Fisch in \cite{CamposHamiltonianFormulationOf}.

Shortly after the introduction of the complex Ashtekar variables, Jacobson generalised the construction to $d=4$, $N=1$ Supergravity \cite{JacobsonNewVariablesFor}. In the following, different authors including F\"ul\"op \cite{FulopAboutASuperAshtekar}, Gorobey and Lukyanenko \cite{GorobeyTheAshtekarComplex}, as well as Matschull \cite{MatschullAboutLoopStates}, explored the subject further. Armand-Ugon, Gambini, Obr\'egon and Pullin \cite{Armand-UgonTowardsALoop} formulated the theory in terms of a GSU$(2)$ connection and thus unified bosonic and fermionic variables in a single connection. Building on these works, Ling and Smolin published a series of papers on the subject  \cite{LingSupersymmetricSpinNetworks, LingHolographicFormulationOf, LingElevenDimensionalSupergravity}, where, among other topics, supersymmetric spin networks coming from the GSU$(2)$ connection were studied in detail. 
In the above works, complex Ashtekar variables are employed for which the methods developed in  \cite{AshtekarRepresentationsOfThe, AshtekarRepresentationTheoryOf, AshtekarDifferentialGeometryOn, AshtekarProjectiveTechniquesAnd, MarolfOnTheSupport, AshtekarQuantizationOfDiffeomorphism} are not available. Also, the Ashtekar variables are restricted to four spacetime dimensions and thus not applicable to higher dimensional Supergravities. Aiming at a unification of String theory and LQG, Smolin explored non-perturbative formulations of certain parts of eleven dimensional Supergravity \cite{SmolinChernSimonsTheory, SmolinAQuantizationOf}. The generalisation of the Loop Quantum Gravity methods to antisymmetric tensors was considered by Arias, di Bartolo, Fustero, Gambini, and Trias \cite{AriasSecondQuantizationOf}. The full canonical analysis of $d=4$, $N=1$ Supergravity using real Ashtekar-Barbero variables was first performed by Sawaguchi \cite{SawaguchiCanonicalFormalismOf}. Kaul and Sengupta \cite{SenguptaCanonicalSupergravityWith} considered a Lagrangian derivation of this formulation using the Nieh-Yan topological density. 

An attempt to construct Ashtekar-type variables for $d=11$ Supergravity has already been made by Melosch and Nicolai using an $\text{SO}(1,2) \times \text{SO}(16)$ invariant reformulation of the original CJS theory \cite{MeloschNewCanonicalVariables}. In this formulation 
the connection is not Poisson commuting thus forbidding LQG techniques. In a paper on canonical Supergravity in $2+1$ dimensions \cite{MatschullCanonicalQuantumSupergravity}, Matschull and Nicolai discovered a similar noncommutativity property which they avoided by adding a purely imaginary fermionic bilinear to the connection, leading to a complexified gauge group. As was observed in \cite{ThiemannKinematicalHilbertSpaces}, this problem can be avoided by using half-densitised fermions as canonical variables.

The general picture emerging is  that the canonical decomposition  $S = \int dt \, (p \dot{q} - H)$ in the time gauge leads to 
\begin{eqnarray} 
S = \int_\mathcal{\sigma} d^Dx \, \int dt \, \biggl( && \dot{E}^a_i K_a^i + i \sqrt{q} \bar{\psi}_a \gamma^{a \perp b} \dot{\psi}_b + \text{tensors} +\text{vectors}+ \text{spin }1/2 +\text{scalars} \nonumber \\
 && - N \mathcal{H} - N^a \mathcal{H}_a - \lambda_{ij} G^{ij} - \bar{\psi}_t \mathcal{S} -\text{tensor constraints}\biggr) \text{,}
\end{eqnarray}
where the Hamiltonian constraint $\mathcal{H}$, the spatial diffeomorphism constraint $\mathcal{H}_a$, the Spin$(D)$ Gau{\ss} constraint $G^{ij}$, the supersymmetry constraint $\mathcal{S}$ and the tensor constraints form a first class algebra. $E^a_i$ is the densitised vielbein and $K_a^i$ its canonical momentum. $\psi_a$ denotes the Rarita-Schwinger field with suppressed spinor indices. $N$, $N^a$, $\lambda_{ij}$ and $\bar{\psi}_t$ are Lagrange multipliers for the respective constraints. With tensor constraints we mean constraints acting only on additional tensor fields  such as the three-index photon of $D+1=10+1$ Supergravity. The remaining terms in the first line are kinetic terms appearing in the decomposition of the action. Since we will not deal with them in this paper, we refer to \cite{HenneauxHamiltonianFormulationOf, DiazHamiltonianFormulationOf} for details. 

In order to apply the techniques developed for Loop Quantum Gravity to this system, we have to turn it into a connection formulation in the spirit of the Ashtekar variables. Concerning the purely gravitational part, this has been achieved in \cite{BTTI, BTTII} and extended to the case of spin $1/2$ fermions in \cite{BTTIV}. 
The Rarita-Schwinger field turns out to be more difficult to deal with than the spin $1/2$ fermions. On the one hand, it leads to second class constraints \cite{PilatiTheCanonicalFormulation}, which encode the reality conditions, with a structure which is different from the case of Dirac spinors\footnote{While for Dirac spinors, the second class constraints are of the form $\pi_{\bar{\psi}} \propto \psi$, $\pi_{\psi} \propto \bar{\psi}$, in the Majorana case we obtain an equation of the form $\pi_{\psi} \propto \psi$, where $\pi_{x}$ denotes the momenta conjugate to $x$.}. On the other hand, as the other fermions, it has to be treated as a half-density in order to commute with $K_a^i$\cite{DeserHamiltonianFormulationOf}.

Apart from the conventional canonical analysis, where time and space are treated differently, there exists a covariant canonical formalism treating space and time on an equal footing \cite{DaddaCovariantCanonicalFormalism}. It has been applied to vielbein gravity \cite{NelsonCovariantCanonicalFormalism}, $d=4$, $N=1$ Supergravity \cite{LerdaCovariantCanonicalFormalism},  $d=5$ Supergravity and higher dimensional pure gravity \cite{FoussatsCanonicalCovariantFormalism, FoussatsHamiltonianFormalismForHigher} and $d=10$, $N=1$ Supergravity coupled to supersymmetric Yang-Mills theory \cite{FoussatsHamiltonianFormalismFor, FoussatsSecondOrderHamiltonian}.  The relation of the covariant canonical formalism and the conventional canonical analysis is discussed in \cite{FoussatsAlgebraOfConstraints} using the example of four dimensional Supergravity coupled to supersymmetric Yang-Mills theory.

\subsection{Canonical Supergravity in the Time Gauge}

We will illustrate the $3+1$ split of $N=1$ Supergravity in first order formulation as performed by Sawaguchi \cite{SawaguchiCanonicalFormalismOf} in order to give the reader a feeling for what is happening during the canonical decomposition. The resulting picture generalises to all
dimensions. The symplectic potential derived in this context is exemplary for the Supergravity theories of our interest and we will continue with the general treatment in the next section. We remark that in $3+1$ spacetime dimensions, the relations $C^T = -C$ and $C\gamma^{I}C^{-1} = - (\gamma^{I})^T$ hold, where $C$ denotes the charge conjugation matrix. 

The action for $3+1$, $N=1$ first order Supergravity is given by
\be
S = \int_{\mathcal{M}} d^{4}X \left( \frac{s}{2} e e^{\mu I} e^{\nu J} F_{\mu \nu IJ} (A)+i s ~ e \bar{\psi}_{\mu} \gamma^{\mu \rho \sigma} \nabla_{\rho}(A) \psi_{\sigma}  \right) \text{.}
\ee
Using the conventions introduced above and $\gamma^{\mu \rho \sigma} = \gamma^{IJK} e^{\mu}_I e^{\rho}_J e^{\sigma}_K$, one can explicitly check that the action is real. The $3+1$ decomposition is done like in the previous papers, and  the notation used can be found there (if not above). We obtain
\ba
S &=& \int_{\mathcal{M}} d^{4}X \left( \frac{s}{2} e e^{\mu I} e^{\nu J} F_{\mu \nu IJ} (A)+ s \,i\, e \bar{\psi}_{\mu} \gamma^{\mu \rho \sigma} \nabla_{\rho}(A) \psi_{\sigma}  \right) \nonumber \\
	&=& \int_{\mathbb{R}} dt \int_{\sigma} d^{3}x  \left[  \frac{1}{2} \pi^{aIJ} \mathcal{L}_T A_{aIJ} - i \sqrt{q} \; \bar{\psi}_{a} \gamma^{\perp ab} \mathcal{L}_T \psi_b   - \utilde{N}\left(\mathcal{H}^{\text{grav}} - i \sqrt{q} \; \bar{\psi}_a \gamma^{abc} \nabla_b(A) \psi_c \right)   \right. \nonumber \\
	&~& ~~~ \left. -N^a \left(  \mathcal{H}_a^{\text{grav}} + 3  i \sqrt{q} \bar{\psi}_{[a} \gamma^{\perp bc} \nabla_{b}(A) \psi_{c]}  \right) + \frac{1}{2} A_{t IJ} \left( G^{IJ}_{\text{grav}} -  i \sqrt{q} \bar{\psi}_a \gamma^{\perp ab} [i\Sigma^{IJ}] \psi_b \right) \right. \nonumber \\
	&~& ~~~\left. -i \bar{\psi}_t \left(  \sqrt{q} \gamma^{\perp ab} \nabla_b(A) \psi_a +  \sqrt{q} \nabla_b(A)(\gamma^{\perp ab}\psi_a )\right) \right] \text{.}
\ea
From there, one can read off the constraints $\mathcal{H}$, $\mathcal{H}_a$, $G^{IJ}$ and $\mathcal{S}$. We will choose time gauge $n^I = \delta^{I}_0$ at this point to simplify the further discussion. For the symplectic potential, we find
\ba
 &~& \int_{\mathbb{R}} dt \int_{\sigma} d^{3}x  \left( \frac{1}{2} \pi^{aIJ} \mathcal{L}_T A_{aIJ} - i \sqrt{q} \bar{\psi}_{a} \gamma^{\perp ab} \mathcal{L}_T \psi_b \right) \nonumber \\
 	&\rightarrow& \int_{\mathbb{R}} dt \int_{\sigma} d^{3}x \left( \dot E^{ai}  K_{ai} - i \phi^{\dagger}_{a} \gamma^{ab} \dot \phi_b  \right) \nonumber \\
	&=& \int_{\mathbb{R}} dt \int_{\sigma} d^{3}x \left( \dot E^{ai}  K_{ai} - i \phi^{\dagger}_{i} \gamma^{ij} \left[ \dot \phi_j -  \dot {(E^{b}_{j})} E_b^k \phi_k \right]  \right) \nonumber \\
	&=& \int_{\mathbb{R}} dt \int_{\sigma} d^{3}x \left( \dot E^{ai}  K_{ai} - \pi^j \left[ \dot \phi_j -  \dot {(E^{b}_{j})} E_b^k \phi_k \right]  \right) \nonumber \\
	&=& \int_{\mathbb{R}} dt \int_{\sigma} d^{3}x \left( \dot E^{ai}  {(K_{ai} + \pi_i E_a^j \phi_j)} - \pi^j \dot \phi_j \right) \nonumber \\
	&=& \int_{\mathbb{R}} dt \int_{\sigma} d^{3}x \left( \dot E^{ai}  K'_{ai} - \pi^j \dot \phi_j \right) \text{,} \label{eq:SymplecticPotential}
\ea
where we successively defined 
\be \label{eq:definitions1}
\phi_a := \sqrt[4]{q} \psi_a ~~\text{,}~~ \phi_i := \frac{1}{\sqrt{q}} E^{a}_i \phi_a ~~\text{,}~~ \pi^i := i \phi_j^{\dagger} \gamma^{ji} ~~\text{and}~~ K'_{ai} := K_{ai}+ \pi_i E_a^j \phi_j~~\text{.}
\ee
 In the second line, we chose time gauge and half-densities as fermionic variables \cite{ThiemannKinematicalHilbertSpaces}. Then, we transformed the spatial index of the fermions into an internal one using the vielbein, but preserving the fermionic density weight \cite{DeserHamiltonianFormulationOf}. This second transformation also affects the extrinsic curvature and we have to define a new variable $K'_{ai}$.
The Gau{\ss} constraint becomes under these changes of variables
\ba
G^{ij} &=& 2 K_a^{[i} E^{a|j]} - \pi^k \left[ i \Sigma^{ij}\right] \phi_k  \nonumber \\
	 &=& 2 \left(K'_a\m^{[i} - \pi^{[i} E_a^k \phi_k \right) E^{a|j]} - \pi^k \left[ i \Sigma^{ij}\right] \phi_k  \nonumber \\
	 &=& 2 K'_a\m^{[i} E^{a|j]} - 2 \pi^{[i} \phi^{j]} - \pi^k \left[ i \Sigma^{ij}\right] \phi_k  \text{.}
\ea
The generator of spatial diffeomorphisms $\tilde{\mathcal{H}}_a$ is given by the following linear combination of constraints
\be
\tilde{\mathcal{H}}_a := \mathcal{H}_a + \frac{1}{2} A_{aij} G^{ij} + i\, \bar{\psi}_a \mathcal{S} \text{.}
\ee
It becomes
\ba
\tilde{\mathcal{H}}_a &=& E^{bj} \partial_a K_{bj} - \partial_b \left( E^{bj} K_{aj} \right) - \pi^b \partial_a \phi_b + \partial_b \left( \pi^b \phi_a \right)  \nonumber \\
	&=& E^{bj} \partial_a \left(K'_{bj} + \pi_j E_b^k \phi_k\right) - \partial_b \left( E^{bj} \left(K'_{aj}+  \pi_j E_a^k \phi_k \right) \right) \nonumber \\
	& & - \frac{1}{\sqrt[4]{q}} \pi_i E^{bi} \partial_a \left( \sqrt[4]{q} \phi^j E_{bj} \right) + \partial_b \left( \pi_i E^{bi} \phi^j E_{aj} \right)  \nonumber \\
	&=& E^{bj} \partial_a K'_{bj} - \partial_b \left( E^{bj} K'_{aj} \right) + \sqrt[4]{q} \partial_a \left( \frac{1}{\sqrt[4]{q}}\pi_i \right) \phi^i   \nonumber \\
	&=& E^{bj} \partial_a K'_{bj} - \partial_b \left( E^{bj} K'_{aj} \right) + \frac{1}{2} \partial_a \left(\pi_i \right) \phi^i  -  \frac{1}{2}  \pi_i \partial_a \phi^i 	\text{.}
\ea
For the last step, note that $\pi^i \phi_i = 0$. Thus, these constraints exactly change as one would expect under the performed change of variables. The other constraints also can be rewritten in terms of the new variables, but this is less instructive and their explicit form is not important for what follows. We only want to remark that they depend on the contorsion $K_{aij}$, which is not dynamical and has to be solved for in terms of $\phi_i$. This can be done explicitly.

\section{Phase Space Extension}

In this section we focus on the symplectic structure of the Rarita-Schwinger sector.
In the time gauge this is a SO$(D)$ theory which is the subject of the first part. 
In the second part we will perform a phase space extension to a SO$(D+1)$ theory
where special attention must be paid to the reality conditions.

\subsection{Symplectic Structure in the SO$(D)$ Theory}

The $3+1$ split described above generalises directly to higher dimensions. We will always impose the time gauge $n^I = \delta^I_0$ prior to the $D+1$ split and restrict to dimensions where a Majorana representation of the $\gamma$-matrices exists, which we will use. This allows us to set $C = \gamma^0$ which simplifies the following analysis. The generic terms important for this paper appearing in Supergravity theories are 
\be
S_{grav. + RS} = \int_{\mathcal{M}} d^{D+1}X \left( \frac{s}{2} e e^{\mu I} e^{\nu J} F_{\mu \nu IJ} (A)+ i s ~ e \bar{\psi}_{\mu} \gamma^{\mu \rho \sigma} \nabla_{\rho}(A) \psi_{\sigma}  \right) 
\ee
in case of a first order formulation and analogous terms for a second order formulation. This difference in defining the theory will not be important in what follows, since as demonstrated above for the $3+1$ dimensional case, the symplectic potential of these actions in the time gauge turns out to be 
\ba
 && \int_{\mathbb{R}} dt \int_{\sigma} d^{D}x  \left( - E^{ai} \mathcal{L}_T K_{ai} - i \sqrt{q} \bar{\psi}_{a} \gamma^{\perp ab} \mathcal{L}_T \psi_b \right) \nonumber \\
&=& \int_{\mathbb{R}} dt \int_{\sigma} d^{D}x \left( \dot E^{ai}  K'_{ai} - \pi^j \dot \phi_j \right) \text{,} \label{eq:SymplecticPotentialD}
\ea
where we used the same definitions as in (\ref{eq:definitions1}). From (\ref{eq:SymplecticPotentialD}) we can read off the non-vanishing Poisson brackets\footnote{More precisely we should call them Poisson anti-brackets which are symmetric under exchange of the arguments and which are to be quantised by anti commutators. We will call them Poisson brackets anyway for notational simplicity in what follows with the usual rules for the interplay between the Poisson brackets for integral and half-integral spin respectively. See e.g. \cite{HenneauxQuantizationOfGauge, GitmanQuantizationOfFields} for an account.} 
\be
\left\{E^{ai}, {K'}_{bj}\right\} = \delta^a_b \delta^i_j ~~\text{and}~~ \left\{\phi_{i}^{\alpha}, \pi^j_{\beta}\right\} = -\delta^{\alpha}_{\beta} \delta_i^j \text{.}
\ee
Additionally, we have the following second class constraints and reality conditions
\be
\Omega^i := \pi^i + i \phi_{j}^{T} C \gamma^0 \gamma^{ji} = 0 ~~\text{and}~~ \phi_{i}^{\dagger} = -\phi_{i}^{T}C\gamma^0 \text{.} \label{eq:RealityConditions}
\ee
In order to be able to introduce a connection variable along the lines of \cite{BTTI}, we need to enlarge the internal space, i.e. replacing the gauge group SO$(D)$ by either SO$(1,D)$ or SO$(D+1)$. In view of subsequent quantisation, SO$(D+1)$ is favoured because of its compactness and will be our choice in the following. This enlargement can be done consistently if also additional spinorial degrees of freedom are added as well as additional constraints which remove the newly introduced fermions. Finally, the extension has to be consistent with the reality conditions. All this turns out to be rather hard to achieve, and the final version of the theory looks rather different from what a ``first guess" might have been. To motivate it, we will review the whole process of finding the theory, showing where the straight-forward ideas lead to dead ends, and how they can be modified to arrive at a consistent theory. We will only discuss the fermionic variables, the gravitational part is treated in the appendix.

Before we enlarge the internal space, we will get rid of the second class constraints. To this end, we calculate the Dirac matrix
\be
C^{ij} = \left\{\Omega^i, \Omega^j\right\} = - 2 i C\gamma^0\gamma^{ij} ~~\text{,}~~ (C^{-1})_{ij} =- \gamma^0 \frac{i}{2(D-1)} \left( \left(2-D\right)\eta_{ij} + \gamma_{ij} \right) C^{-1}\text{,}
\ee
and thus find for the Dirac bracket
\be
\left\{\phi_i, \phi_j \right\}_{DB} = -\left\{\phi_i, \Omega^k \right\} (C^{-1})_{kl} \left\{\Omega^l ,\phi_j\right\} = - (C^{-1})_{ij} \text{.} \label{eq:DiracBracket}
\ee
To simplify the subsequent discussion, in the following we will consider real representations of the Dirac matrices only, which implies $C = \gamma^0$. Then the above equations read
\be
C^{ij} = 2i \gamma^{ij} ~~\text{,}~~ (C^{-1})_{ij} = -\frac{i}{2(D-1)} \left( \left(2-D\right)\eta_{ij} + \gamma_{ij} \right) ~~\text{,}~~ \left\{\phi_i, \phi_j \right\}_{DB} = -(C^{-1})_{ij} \text{.}
\ee

Now we can either (a) try to enlarge the internal space and afterwards choose new variables which have simpler brackets, or (b) we simplify the Dirac bracket before enlarging the internal space. (a) immediately leads to problems. The symmetry of the Poisson brackets $\left\{\phi^{\alpha}_I, \phi_J^{\beta}\right\} \propto (\tilde C^{-1})_{IJ}^{\alpha \beta}$ implies that matrix $\tilde{C}^{-1}$ is symmetric under the exchange of $(I, \alpha) \leftrightarrow (J,\beta)$. The naive extension $(C^{-1})_{IJ} = -\frac{i}{2(D-1)} \left( \left(2-D\right)\eta_{IJ} + \gamma_{IJ} \right)$ however does not have this symmetry. Its symmetric part $\tilde{C}^{-1} + (\tilde{C}^{-1})^T$ is not invertible. Of course, one can extend $C^{-1}$ in different, more ``unnatural" ways, e.g. containing terms like $\gamma_J^T \gamma_I$ etc. and ``cure" this problem for a moment, but also the Gau{\ss} constraint will be problematic. The SO$(D)$ constraint contains $C^{ij}$ (since we used $\pi^{i} =- \frac{1}{2} \phi_j^T C^{ji}$) and this matrix should also be replaced by some $\hat C^{IJ}$, such that $\phi_I$ transforms covariantly and $G^{IJ}$ reduces correctly to $G^{ij}$ if we choose time gauge and solve its boost part. This implies restrictions on $\hat{C}$ and further restrictions on $\tilde{C}^{-1}$. We did not succeed in finding matrices which fulfil all these requirements. In the following, we therefore will follow the second route (b) and simplify the Dirac brackets before doing the enlargement of the internal space.\\
\\
There are several possible ways how to simplify the Dirac brackets:
\begin{enumerate}
\item Note that the matrix $C^{-1}$ on the right hand side of the Dirac brackets is imaginary and symmetric, hence there always exists a real, orthogonal matrix $O^i\m_j$ such that under the change of variables $\phi^i \rightarrow \phi'\m^i := O^i\m_j\phi^j$ the brackets becomes $i$ times a real diagonal matrix. However, now the new fundamental degrees of freedom $\phi'\m^i$ in general do not transform nicely under SO$(D)$ gauge transformations, only $(O^{-1})^i\m_j \phi'\m^j$ do. More severely, it is unclear how the extension $O^i\m_j \rightarrow O^I\m_J$ should be done.
\item To assure that the fundamental degrees of freedom still transform nicely under SO$(D)$ transformations, we can use the Ansatz $\phi'\m^i := M^{ij} \phi_j$ with $M^{ij} := (\alpha \delta^{ij} \mathbb{1}+ \beta \Sigma^{ij})$. Matrices of this form are in general invertible (cf. point 3. below for two exceptions) and, since they are constructed from intertwining matrices, $\phi'\m^i$ will transform nicely under gauge transformations. Moreover, now there is a chance to generalise the matrix to one dimension higher. For the Dirac brackets to become diagonal, $\alpha$ and $\beta$ have to be determined by solving $MC^{-1}M^T = i \mathbb{1}$. The problem is that there is no solution for both parameters being real, at least one is necessarily complex. More general Ans\"atze for $M^{ij}$ (e.g. involving $\gamma_{\text{five}}$ in even dimensions) share the same problem. Thus we exchanged the problem of complicated brackets with complicated reality conditions, which again are hard to quantise.
\item The third route, which will lead to the consistent theory, in the end implies the introduction of additional fermionic degrees of freedom already before enlargement of the internal space.
Given the difficulties just mentioned, the optimal approach in the desire to simplify 
the Poisson brackets is to find orthogonal projections onto subspaces of the real Gra{\ss}mann 
vector space which are built from $\delta_{ij}\mathbb{1}$ and $\Sigma_{ij}$ such that the 
symplectic structure becomes block diagonal on those subspaces. One can then define 
simple Poisson brackets and add the projection constraints as secondary constraints which
leads to corresponding Dirac brackets which will be proportional to those projectors.
As we will see, the fact that these are projectors makes it possible to find a Hilbert space 
representation of the corresponding Dirac bracket.\\ 
\\ 
We define in any dimension $D$ 
\ba
\mathbb{P}^{ij}_{\alpha \beta} &:=& \eta^{ij} \delta_{\alpha \beta} - \frac{1}{D} (\gamma^i \gamma^j)_{\alpha \beta} = \frac{D-1}{D} \eta^{ij} \delta_{\alpha \beta} - \frac{2i}{D} \Sigma^{ij}_{\alpha \beta} \text{,} \\
\mathbb{Q}^{ij}_{\alpha \beta} &:=& \frac{1}{D} (\gamma^i \gamma^j)_{\alpha \beta} = \frac{1}{D} \eta^{ij} \delta_{\alpha \beta} + \frac{2i}{D} \Sigma^{ij}_{\alpha \beta} \text{.}
\ea
Those matrices are both real (we are using Majorana representations) and built from intertwiners, but they are not invertible. It is easy to check that
\be
	\mathbb{P}^{ij}_{\alpha \beta} \mathbb{Q}^{\beta \gamma}_{j k} = 0 \text{,}~~~\mathbb{P}^{ij}_{\alpha \beta} \mathbb{P}^{\beta \gamma}_{j k} = \mathbb{P}^{i \gamma}_{\alpha k} \text{,}~~~ \mathbb{Q}^{ij}_{\alpha \beta} \mathbb{Q}^{\beta \gamma}_{j k} =  \mathbb{Q}^{i \gamma}_{\alpha k}, ~~~\text{and}~~~\mathbb{P} + \mathbb{Q} = \mathbb{1} \eta \text{,}
\ee 
i.e. the above equations define projectors. By construction, $\mathbb{P}$ projects on ``trace-free" components w.r.t. $\gamma_i$, i.e. $\mathbb{P}^{ij}_{\alpha \beta} \gamma^{\beta}_j = 0 = \gamma_{i}^{\alpha} \mathbb{P}^{ij}_{\alpha \beta}$. Using these projectors, we can decompose the Rarita-Schwinger field as follows
\be
	\phi_i = \mathbb{P}_{ij}\phi^j + \mathbb{Q}_{ij}\phi^j =:  \rho_i + \frac{1}{D} \gamma_i \sigma\text{,} \label{eq:DecompositionFermions}
\ee
with $\rho_i := \mathbb{P}_{ij}\phi^j$ and $\sigma := \gamma^i \phi_i$\footnote{When considering the free Rarita-Schwinger action, this decomposition also appears to isolate the physical degrees of freedom, cf. e.g. \cite{DeserHamiltonianFormulationOf}. The ``trace part" $\sigma$ is unphysical for the free field.}. Using the reality conditions (\ref{eq:RealityConditions}) for $\phi_i$, we find
\be
\bar{\rho}_i = \rho_i^T C ~~\text{and}~~\bar{\sigma} = \sigma^T C \text{.} \label{eq:RealityConditions2}
\ee
Moreover, using
\be
	\gamma^{ij} = - \mathbb{P}^{ij} + (D-1) \mathbb{Q}^{ij} \text{,}
\ee
the symplectic potential becomes
\ba
-\pi^i \dot \phi_i &=& - i \phi^{\dagger}_j \gamma^{ji} \dot \phi_i \nonumber \\
					&=& - i\phi^{\dagger}_j \left( - \mathbb{P}^{ji} +(D-1) \mathbb{Q}^{ji} \right) \dot \phi_i \nonumber \\
					&=& - i\phi^{\dagger}_j \left( - \mathbb{P}^{j}\m_k \mathbb{P}^{ki} +(D-1) \mathbb{Q}^{j}\m_k \mathbb{Q}^{ki} \right) \dot \phi_i \nonumber \\
					&=&  i\left(  \mathbb{P}_k\m^j \phi_j \right)^{\dagger} \dot{\left(\mathbb{P}^{ki} \phi_i\right)} - i(D-1)\left( \mathbb{Q}_k\m^{j}\phi_j\right)^{\dagger} \dot{\left(\mathbb{Q}^{ki} \phi_i\right)} \nonumber \\
					&=& i\rho^{\dagger}_i  \dot \rho^i - i\frac{D-1}{D} \sigma^{\dagger} \dot \sigma \nonumber \\
					&=& - i\rho^{T}_i C \gamma^0  \dot \rho^i + i\frac{D-1}{D} \sigma^T C \gamma^0 \dot \sigma \nonumber \\
					&=& i\rho^{T}_i \dot \rho^i - i\frac{D-1}{D} \sigma^T \dot \sigma \text{,}
\ea
where in the second to last line we used the reality conditions (\ref{eq:RealityConditions2}) and in the last line we restricted to a real representation, $C = \gamma^0$.
\end{enumerate}
This motivates the definition of the brackets
\be
\left\{ \rho_j,  \rho^i \right\} = -\frac{i}{2} \mathbb{1} \delta^i_j ~~\text{and}~~ \left\{\sigma, \sigma \right\} = i \frac{D}{2(D-1)} \mathbb{1} \text{,} \label{eq:PoissonBrackets2}
\ee
together with the reality conditions $\rho^*_i = \rho_i$, $\sigma^* = \sigma$ (cf. (\ref{eq:RealityConditions2})) and additionally introduced constraints to account for the superfluous fermionic degrees of freedom,
\be
\Lambda_{\alpha} := \gamma^i_{\alpha \beta} \rho^{\beta}_i \approx 0 \text{.}
\ee
We need to check that the extension is valid, i.e. that the Poisson brackets of the $\phi_i$, considered as functions on the extended phase space, are equal to the Dirac brackets (\ref{eq:DiracBracket}) of the system before we did the extension. Using $\phi_i = \mathbb{P}_{ij}\rho^j + \frac{1}{D} \gamma_i \sigma$ (cf. \ref{eq:DecompositionFermions}) and the Poisson brackets (\ref{eq:PoissonBrackets2}), this can be checked explicitly (this calculation shows why the factors of $\frac{1}{2}$ in (\ref{eq:PoissonBrackets2}) are needed). Using this, we can express the constraints $\mathcal{H}$ and $\mathcal{S}$ in terms of the new variables in the obvious way and know that their algebra is unchanged. In particular, since the projectors are built from intertwiners, we find for the fermionic part of the Gau{\ss} constraint
\be
G^{ij} = ... + \left( i \rho^{kT}\right) \left[ 2\eta^{[i}_k\eta_l^{j]} + i \Sigma^{ij} \eta_{kl}\right] \rho^l + \left(- i \frac{D-1}{D}\sigma^T \right) \left[i\Sigma^{ij}\right]\sigma \text{,}
\ee
which allows for an easy generalisation to SO$(D+1)$ or SO$(1,D)$ as a gauge group. Furthermore, since $\rho^i$ in the other constraints only appears in the combination $\mathbb{P}_{ij}\rho^j$, they automatically Poisson commute with $\Lambda_{\alpha}$. 

Note that if we now would calculate the Dirac bracket, we would get $\{\rho_i,\rho_j\}_{DB} = -\frac{i}{2} \mathbb{P}_{ij}$, which again is non-trivial. Instead, we directly enlarge the phase space from $\{\rho^i, \sigma\}$ to $\{\rho^I, \sigma\}$, with, as a first guess, the brackets $\left\{ \rho_I, \rho_J\right\} = -\frac{i}{2} \eta_{IJ} \mathbb{1}$, $\left\{\sigma,\sigma\right\} = i \frac{D}{2(D-1)} \mathbb{1}$, the reality conditions $\rho_I^* = \rho_I$, $\sigma^* = \sigma$ and the constraints
\be
N^I \rho_I \approx 0 ~~\text{and}~~ \gamma^I \rho_I \approx 0 \text{.} \label{eq:NewConstraints}
\ee
Unfortunately, this immediately leads to an inconsistency in the case of the compact gauge group SO$(D+1)$, since for our choice of Dirac matrices, $\gamma^0$ necessarily is complex in the Euclidean case. Therefore, the reality conditions again are not SO$(D+1)$ covariant and the constraints (\ref{eq:NewConstraints}) only are consistent in the time gauge $N^I = \delta^I_0$\footnote{$\gamma^I \rho_I \approx 0$ is a complex constraint and thus equal to two real constraints. Only in time gauge, its imaginary part is already solved by demanding $N^I \rho_I \approx 0 $.}. With a more elaborate choice of reality condition it is possible to define a consistent theory, which will be the subject of the next section.

\subsection{SO$(D+1)$ Gauge Supergravity Theory}

As we just have seen, the remaining obstacle on our road of extending the internal gauge 
group from SO$(D)$ to SO$(D+1)$ is that the real vector space $V$ of real SO$(1,D)$ Majorana spinors is
 not preserved under SO$(D+1)$ whose spinor representations are necessarily on complex 
 vector spaces. Let $V_{\mathbb{C}}$ be the complexification of $V$. Now SO$(D+1)$ acts on 
 $V_{\mathbb{C}}$ but the theory we started from is not $V_{\mathbb{C}}$ but rather 
 the SO$(D+1)$ orbit of $V$. This is the real vector subspace 
 \be \label{1}
 V_{\mathbb{R}}=\{\theta\in V_{\mathbb{C}};\;\;\exists \; \rho\in V,\;g\in \text{SO}(D+1)\;\;\ni
 \;\;\theta=g\cdot \rho\} \text{,}
 \ee
 where $g\cdot$ denotes the respective representation of SO$(D+1)$. This
 defines a reality structure on $V_{\mathbb{C}}$ that is 
 $V_{\mathbb{C}}=V_{\mathbb{R}}\oplus i V_{\mathbb{R}}$. The mathematical problem 
 left is therefore to add the reality condition that we are dealing with $V_{\mathbb{R}}$ rather
 than $V_{\mathbb{C}}$. 
 
 In order to implement this, recall that any $g\in \text{SO}(D+1)$ can be written as $g=BR$ where
 $B$ is a ``Euclidean boost'' in the $0j$ planes and $R$ a rotation that preserves the internal vector 
 $n^I_0:=\delta^I_0$. The spinor representation of $R$ just needs $\gamma_j$ which is 
 real valued. It follows that (\ref{1}) can be replaced by
  \be \label{2}
 V_{\mathbb{R}}=\{\theta\in V_{\mathbb{C}};\;\;\exists \; \rho\in V,\;B\in \text{SO}(D+1)\;\;\ni
 \;\;\theta=B\cdot \rho\} \text{.}
 \ee
The problem boils down to extracting from a given $\theta\in V_{\mathbb{R}}$ the boost
$B$ and the element $\rho\in V$, that is, we need a kind of polar decomposition. 
If $V_{\mathbb{C}}$ would be just a vector subspace of some $\mathbb{C}^n$ we could do this by standard
methods. But this involves squaring of and dividing by complex numbers and these 
operations are ill defined for our $V_{\mathbb{C}}$ since Gra{\ss}mann numbers are nilpotent. Thus, 
we need to achieve this by different methods.

The natural solution lies in the observation that if we use the linear simplicity constraint
then the $D$ boost parameters can be extracted from the $D$ rotation angles in
the normal $N^I=B_{IJ} n_0^J$ to which we have access because $N$ is part of the 
extended phase space. To be explicit, let $e^{(A)}$ be the standard base of 
$\mathbb{R}^{D+1}$, that is, $e^{(A)}_I=\delta^A_I$. We construct another orthonormal
basis $b^{(A)}$ of $\mathbb{R}^{D+1}$ as follows:\\
Let $b^{(0)}:=N$ and 
\be \label{3}
b^{(0)}_0=\sin(\phi_1)..\sin(\phi_D),\;\;
b^{(0)}_j=\sin(\phi_1)..\sin(\phi_{D-j})\cos(\phi_{D+1-j});\;j=1..D \text{,}
\ee
with $\phi_1,..\phi_{D-1}\in [0,\pi]$ and $\phi_{D}\in [0,2\pi]$ modulo usual 
identifications and singularities of polar coordinates. Define 
\be \label{4}
b^{(j)}_I=\frac{\partial b^{(0)}_I/\partial \phi_j}{||\partial b^{(0)}/\partial \phi_j||} \text{,}
\ee
where the denominator denotes the Euclidean norm of the numerator. Then it maybe checked 
by straightforward computation that 
\be \label{5}
\delta^{IJ}\; b^{(A)}_I\; b^{(B)}_J=\delta^{AB}  \text{.}
\ee
We consider now the SO$(D+1)$ matrix
\be \label{6}
(A(N)^{-1})_{IJ} :=\sum_{A=0}^D\; b^{(A)}_I\; e^{(A)}_J \text{,}
\ee
which has the property that $A(N)^{-1}\cdot e^{(0)}=N$. 

Now starting from the time gauge, $g\in \text{SO}(D+1)$ acts on $V$ and produces
$N=g\cdot e^{(0)}$ and $\theta=g\cdot \rho$. We decompose $g=A(N)^{-1} R(N)$ where 
$A(N)^{-1}$ is the boost defined above and $R\cdot e^{(0)}=e^{(0)}$ is a rotation preserving 
$e^{(0)}$. It follows that we may parametrise any pair $(N,\theta)$ with $||N||=1$ and 
$\theta\in V_{\mathbb{R}}$ as  $A(N)^{-1}\cdot (e^{(0)}, \rho)$ where $\rho\in V$. 
We need to investigate how SO$(D+1)$ acts on this parametrisation. On the one 
hand we have 
\be \label{7}
[g\;A(N)^{-1}]_{IJ}=\sum_A\; (g b^{(A)})_I\; e^{(A)}_J  \text{.}
\ee
On the other hand we can construct $A(g \cdot N)^{-1}$ by following the above procedure,
that is, computing the polar coordinates $\theta_{g j}$ of $g\cdot N$ and defining 
the $b^{(A)}_j(g\cdot N)$ via the derivatives with respect to the $\theta_{g j}$. The common element of both bases is $g\cdot N=g\cdot b^{(0)}$. Therefore, there exists 
an element $R(g,N)\in \text{SO}(D)$ such that 
\be \label{8}
g\cdot b^{j}(N)=R_{kj}(g,N) b^{(k)}(g\cdot N) \text{,}
\ee 
or with $R_{00}=1,\; R_{0i}=R_{i0}=0$
\be \label{9}
 g\cdot b^{A}(N)=R_{BA}(g,N) b^{(B)}(g\cdot N)
 \ee
 defines a rotation in SO$(D+1)$ preserving $e^{(0)}$. Putting these findings together we 
 obtain
 \ba \label{10}
 && [g\cdot A(N)^{-1}]_{IJ}
 =\sum_{A,B}\;  R_{BA}(g,N) \; b^{(B)}_I(g\cdot N)\;e^{(A)}_J
 =\sum_{A}\;  R_{AJ}(g,N) \; b^{(A)}_I(g\cdot N)
 \nonumber\\
& =& \sum_{A}\;  R_{KJ}(g,N) \; b^{(A)}_I(g\cdot N) \; \delta^{(A)}_K  
=[A(g\cdot N)^{-1} R(g,N)]_{IJ}  \text{.}
\ea
~\\
Hence the matrix $A(N)^{-1}$ plays the role of a filter in the sense that the action of SO$(D+1)$
on $A(N)^{-1} \cdot \rho$ can be absorbed into the matrix $A^{-1}$ parametrised by $g\cdot N$ modulo a rotation that preserves $V$ and thus 
altogether the decomposition of  $V_{\mathbb{R}}=\{A(N)^{-1}\cdot V;\;||N||=1\}$ is preserved with 
the expected covariant
action of SO$(D+1)$ on $N$.   
It therefore makes sense to impose the reality condition that $A(N)\, \theta$ is a 
real spinor. In the subsequent construction, this idea will be implemented together with 
an extension of the phase space $\rho_j\to \rho_I$ subject to the constraint $N^I \rho_I=0$.
All these constraints and the reality conditions are second class and we will show 
explicitly that the symplectic structure reduces to the time gauge theory. Despite the
fact that we end up with a non trivial
Dirac (anti-) bracket, it can nevertheless be quantised and non trivial Hilbert space representations can be found as we will demonstrate in the next section.\\
\\
We define $A(N) \in \text{SO}(D+1)$ quite generally\footnote{There exist other possible choices apart from the construction using polar coordinates which might be better suited for certain problems. In $D=3$, we can, e.g., construct $A(N)$ as a linear function of the components of $N^I$ by using $A_{0I} = N_I$ and subsequently interchanging the components of $N^I$ with appropriate signs for the remaining columns of $A(N)$.} in the spin 1 representation by the equation
\be
	A^I\m_J N^J = \delta^{I}_0 \text{.}
\ee
It is determined up to SO$(D)$ rotations. From the above equation, it follows that 
\be
A_{0I} = N_I~~\text{and}~~ A_{IJ} \bar{X}^J = \delta^I_i A_{IJ} \bar{X}^J \label{eq:PropertiesOfA}
\ee
for $X^J$ arbitrary. The corresponding rotation on spinors will be denoted by A. This matrix rotates the normal $N^I$ into its time gauge value $\delta^I_0$ without imposing time gauge explicitly, which we will use to circumvent the reality problems of the SO$(D+1)$ theory mentioned above appearing if we do not choose time gauge. We introduce the set of variables $(A_{aIJ}, \pi^{bKL}, N^I, P_J, \rho_I, \rho^*_J, \sigma, \sigma^*)$ together with the following non-vanishing Poisson brackets
\begin{alignat}{3}
\left\{A_{aIJ}(x), \pi^{bKL}(y) \right\} &= 2 \delta_a^b \delta_I^{[K} \delta_J^{L]} \delta^D(x-y) \text{,}&~~~~~\left\{N^I(x), P_J(y) \right\} &=  \delta_J^I \delta^D(x-y)\text{,}~~ \nonumber \\
\left\{\rho_I(x), \rho^*_J(y) \right\} &= -i \eta_{IJ} \mathbb{1} \delta^D(x-y)\text{,}&~~~\left\{\sigma(x), \sigma^*(y) \right\}& =  i \frac{D}{D-1} \mathbb{1} \delta^D(x-y) \text{,} 
\end{alignat}
and the reality conditions
\be
	\chi_I := A \rho_I - (A \rho_I)^* = 0 \text{,}~~~~ \chi := A\sigma - (A\sigma)^* = 0~~\text{,} \label{eq:NewRealityConditions}
\ee
which just say that the fermionic variables are real as soon as the normal $N^I$ gets rotated into time gauge. Notice that before imposing the constraints, $\rho,\theta$ are complex 
Gra{\ss}mann variables and only the Poisson brackets between these and their complex conjugates are non vanishing. The non vanishing brackets between themselves of the previous section will 
be recovered when replacing the above Poisson bracket by the corresponding Dirac
bracket.

Additionally, we want that the variables transform nicely under spatial diffeomorphisms and gauge transformations, thus we add
\ba
G^{IJ} &:=& D_a \pi^{aIJ} + 2P^{[I}N^{J]} + 2i \rho^{\dagger[I}\rho^{J]} + i\rho^{\dagger}_K [i\Sigma^{IJ}] \rho^K - i\left(\frac{D-1}{D} \sigma^{\dagger}\right) [i\Sigma^{IJ}] \sigma + \hdots \\
\tilde{\mathcal{H}}_a &:=& \frac{1}{2} \pi^{bIJ} \partial_a A_{bIJ} - \frac{1}{2}\partial_b\left(\pi^{bIJ}A_{aIJ}\right) + P^I \partial_a N_I \nonumber \\ &\m& - \frac{i}{2} \partial_a(\rho^{\dagger I}) \rho_I + \frac{i}{2} \rho^{\dagger I} \partial_a \rho_I +i \frac{D-1}{2D} \partial_a(\sigma^{\dagger}) \sigma - i \frac{D-1}{2D} \sigma^{\dagger} \partial_a \sigma + \hdots \text{.}
\ea
The old variables are expressed in terms of the new ones by
\ba
	E^{ai} := \zeta A^{iJ} \bar{\eta}_{JK} \pi^{aIK} N_I \text{,}&\m& K_{ai} = \zeta A_{i}\m^I \bar{\eta}_{IK} (A_{aKJ} - \Gamma_{aKJ}(\pi)) N^J \text{,} \nonumber \\
	\rho_i = \frac{1}{2} A_{iJ} \bar{\eta}^{JK} \left(A\rho_K + A^* \rho^*_K\right) \text{,}&\m& \sigma = \frac{1}{2} \left(A\sigma+ A^* \sigma^*\right) \text{,} \label{eq:OldVariablesNew}
\ea
where the bar here means rotational components w.r.t $N^I$, $\bar{\eta}_{IJ} := \eta_{IJ} - \zeta N_I N_J$. To remove unnecessary degrees of freedom, we need the constraints
\ba
	S^a_{I\overline{M}} &:=& \epsilon_{IJKL\overline{M}} N^J \pi^{aKL} \text{,} \nonumber \\
	\mathcal{N} &:=& N^IN_I - \zeta \text{,} \nonumber  \\
	\Lambda &:=& \gamma^{I} A_{IJ} \bar{\eta}^{JK} (A \rho_K + A^*\rho^*_K)  = A \gamma_{J} \bar{\eta}^{JK} (\rho_K + A^{-1}A^*\rho^*_K) \text{,} \nonumber  \\
	\Theta &:=& N^I (A\rho_I + A^*\rho^*_I) \text{,} \label{eq:NewConstraintsFinal}
\ea
together with the Hamilton and supersymmetry constraints, where we replace the old by the new variables as shown above. To prove that this theory is equivalent to Supergravity and can possibly be quantised, we have to answer the following questions:
\begin{itemize}
\item Are the reality conditions (\ref{eq:NewRealityConditions}) consistent? I. e., do they transform under gauge transformations in a sensible way and do they (weakly) Poisson commute with the other constraints?
\item Are the Poisson brackets of the old variables when expressed in terms of the new ones (\ref{eq:OldVariablesNew}) equal to those on the old phase space? Does the constraint algebra close, i.e. do the newly introduced constraints (\ref{eq:NewConstraintsFinal}) fit ``nicely" in the set of the old constraints? If not, do at least the constraints which were of the first class before the enlargement of the gauge group retain this property?
\item Do the constraints, especially the Gau{\ss} and spatial diffeomorphism constraint, reduce correctly?
\item Which Dirac brackets arise from the reality conditions? In view of a later quantisation, can we find variables such that the Dirac brackets become simple?
\end{itemize}
We will answer these questions in the order they were posed above.
\begin{itemize}
\item The orthogonal matrix $A_{IJ}$ is a function of $N^I$ only as we have seen above. We have $A_{0K} = N_K$, but the remaining components of the matrix are complicated functions of the components of the vector $N^I$. Thus, the whole matrix $A_{IJ}$ will have a rather awkward transformation behaviour under the action of $G^{IJ}$. The reality conditions (\ref{eq:NewRealityConditions}) as a whole, however, transform in a ``nice" way under SO$(D+1)$ gauge transformations (we will discuss $\rho_I$ in the following, $\sigma$ can be treated analogously). For $g \in \text{SO}(D+1)$, the reality condition transforms as follows:
\be
A(N)\rho^J = A(N)^* \rho^{J*} \longrightarrow g^{JK} A(g\cdot N) g\rho_K = g^{*JK} A(g\cdot N)^* g^* \rho^*_K \text{.}
\ee
Since $g^{IJ}$ is real, it is sufficient to consider the transformation behaviour of the spinor $A\rho^I$, so we will skip the action on internal indices in the following. Note that every rotation can be split up in a part which leaves $N^I$ invariant and a ``Euclidean boost" changing $N^I$. For the rotations, $A$ is invariant and we find using $A\bar{\gamma}^I A^{-1} = \bar{\eta}^I\m_JA^{-1}_{JK} \gamma^K$ and $\Sigma^{ij*} = -\Sigma^{ij}$
\ba
\delta_{\bar{\Lambda}}A\rho_I &=& i\bar{\Lambda}_{JK} A\bar\Sigma^{JK}\rho_I = i\bar{\Lambda}_{JK} A\bar\Sigma^{JK}A^{-1}A\rho_I = i\bar{\Lambda}^{JK} A^{-1}_{[J|L} A^{-1}_{K]M} \Sigma^{LM} A \rho_I = \nonumber \\ 
&=& iA_{L[J} A_{M|K]} \bar{\Lambda}^{JK}  \Sigma^{LM} A \rho_I = iA_{l[J} A_{m|K]} \bar{\Lambda}^{JK}  \Sigma^{lm} A \rho_I \text{,} \\
\delta_{\bar{\Lambda}}(A\rho_I)^* &=& (i\bar{\Lambda}_{JK} A\bar\Sigma^{JK}\rho_I)^* = (iA_{l[J} A_{m|K]} \bar{\Lambda}^{JK}  \Sigma^{lm} A \rho_I)^* = \nonumber \\
&=& - iA_{l[J} A_{m|K]} \bar{\Lambda}^{JK}  \Sigma^{lm*} A^* \rho^*_I = iA_{l[J} A_{m|K]} \bar{\Lambda}^{JK}  \Sigma^{lm} A^* \rho^*_I \text{.}
\ea
For finite transformations $\bar{g} \in \text{SO}(D)_N$ stabilising $N^I$, we thus have $A\rho_I \rightarrow A \bar{g} \rho_I = g_0 A \rho_I$, where $g_0 \in \text{SO}(D)_0$ stabilises the zeroth component and thus is, with our choice of representation, a real matrix. Hence, reality conditions transform again into reality conditions under rotations. For a boost $b$ the situation is a bit more complicated. Under a boost $A_{IJ}$ will transform intricately, but we know that a) the matrix remains orthogonal by construction, and b) that $A_{0K} = N_K \rightarrow \Lambda_K\m^L N_L = - A_{0L} \Lambda^L\m_K$. The most general transformation compatible with the above is $A_{IJ}\rightarrow (g_0)_{IK} A^{KL}  (\bar{g}^{-1})_{LN} (b^{-1})^N\m_M (\m^b\bar g^{-1})^M\m_J$ where $g_0 \in \text{SO}(D)_0$ is some group element which does not change the zeroth component, $\bar{g} \in \text{SO}(D)_N$ is in the stabiliser of $N^I$ and $\m^b\bar{g} \in \text{SO}(D)_{b\cdot N}$. Since we have SO$(D)_N = b^{-1}\text{SO}(D)_{b \cdot N} b$, we can eliminate $\m^b\bar{g}$ by a redefinition of $\bar{g}$. By definition of a representation, we then also have $A \rightarrow g_0A\bar{g}^{-1} b^{-1}$ and thus 
\be
A \rho_I\rightarrow g_0A\bar{g}^{-1} b^{-1} b \rho_I = g_0 A \bar{g}^{-1} \rho_I = \tilde{g}_0 A \rho^I \text{,}
\ee
where in the last step we used the result we obtained for rotations above. Since $\tilde{g}_0 \in \text{SO}(D)_0$ is real, we see that under a ``Euclidean boost" the reality condition can only get rotated. What remains to be checked is that the reality condition Poisson commutes with all other constraints. It transforms covariantly under spatial diffeomorphisms by inspection and, as we have just proven, it forms a closed algebra with SO$(D+1)$ gauge transformations. Concerning all other constraints, note that they, by construction, depend only on $\Re(A\rho^J)$ (cf. the replacement (\ref{eq:OldVariablesNew}) and the new constraints (\ref{eq:NewConstraintsFinal})), while the reality condition demands that $\Im(A\rho^J)$ vanishes. But real and imaginary parts Poisson commute, which can be checked explicitly,
\be
\big\{ \left(A \rho_I- A^* \rho^*_I\right) , \left(A \rho_J + A^* \rho^*_J\right) \big\} =  -i \eta_{IJ} \left[+ A A^{\dagger} - A^* A^T \right]  = 0\text{.}
\ee
\item The brackets between $E^{ai}$ and $K_{bj}$ have already been shown to yield the right results in \cite{BTTIV}. The only modifications in the case at hand are a) the replacement of $n^I(\pi)$ by $N^I$ and the corresponding replacement of the quadratic by the linear simplicity constraint, which, in fact, simplifies the calculations, and b) the matrix $A_{IJ}$, which does not lead to problems because of its orthogonality. For the fermionic variables, we find using\footnote{Because of orthogonality, we trivially have $A_{IJ} A_K\m^J = \eta_{IK}$. Additionally, $A_{iJ} \bar{\eta}^{J}_{K} = A_{iK}$, which can be seen from $A_{iK} N^{K} = 0$. Therefore, $A_{i}\m^I A_{j}\m^J \bar{\eta}_{IJ} = A_{i}\m^I A_{jI} = \eta_{ij}$.} $A_{i}\m^I A_{j}\m^J \bar{\eta}_{IJ} = \eta_{ij}$ and $A^{\dagger}A = \mathbb{1}$
\ba
	\left\{\rho_i(x), \rho_j(y)\right\} &=& -\frac{i}{4} A_{i}\m^I A_j\m^J \bar{\eta}_I^K \bar{\eta}_J^L \left[ \left\{A\rho_K(x), A^*\rho^*_L(y) \right\}+ \left\{A^*\rho^*_K(x), A\rho_L(y) \right\}\right] \nonumber \\
	&=& -\frac{i}{4} A_{i}\m^I A_{j}\m^J \bar{\eta}_{IJ} \left[  AA^{\dagger} + A^* A^T\right] \delta^D(x-y) \nonumber \\
	&=& -\frac{i}{4} \delta_{ij} \left[  \mathbb{1} + \mathbb{1}^T\right] \delta^D(x-y) \nonumber \\
	&=& - \frac{i}{2} \delta_{ij} \mathbb{1} \delta^D(x-y) \text{,} \\
	\left\{\sigma(x), \sigma(y) \right\} &=& i \frac{D}{2(D-1)}\mathbb{1} \delta^D(x-y) \text{.}
\ea
This automatically implies that the algebra of $\mathcal{H}$ and $\mathcal{S}$ remains unchanged if we replace the old variables by (\ref{eq:OldVariablesNew}). From (\ref{eq:OldVariablesNew}), it is also clear that $\mathcal{H}$ and $\mathcal{S}$ Poisson commute with $S^a_{I\overline{M}}$ and $\mathcal{N}$. By inspection, all constraints transform covariant under spatial diffeomorphisms. More surprisingly, all constraints Poisson commute with $G^{IJ}$. This can be seen quite easily for $G^{IJ}$, $\tilde{\mathcal{H}}_a$, $S^a_{I\overline{M}}$, $\mathcal{N}$ and also for $\Lambda$ and $\Theta$ (note that $A$, $A_{IJ}$ are invertible and that $\left(\rho_I + A^{-1} A^* \rho^*_I \right)$ transforms like $\rho_I$ which can be shown using the methods above). But for $\mathcal{H}$ and $\mathcal{S}$ this is, at first sight, a small miracle, since the replacement rules (\ref{eq:OldVariablesNew}) of all old variables depend on $A(N)$, which is known to transform oddly. But the matrices $A$ are placed such that they, in fact, either $a)$ appear in the combinations $(\rho_I + A^{-1}A^*\rho^*_I)$ or $(\rho^{\dagger}_I + \rho^T_I A^TA)$, which can easily be shown to transform like $\rho^I$ and $\rho^{\dagger}_I$ respectively with the methods above, or $b)$ all cancel out! The general situation is the following: $\rho_i$ is replaced by $\rho_i = A_i\m^J \bar{\eta}_J\m^I A (\rho_I + A^{-1}A^*\rho^*_I)$, $\rho_i^T$ by  $\rho_i^T = A_i\m^J \bar{\eta}_J\m^I (\rho_I^{\dagger} + \rho^{T} A^T A)A^{-1}$, where the expression in brackets transform sensible (cf. above). The free internal indices of $E^{ai}$, $K_{bj}$ and $\rho^k$ are either contracted with each other, then in the replacement the $A_{iJ}$s will cancel because of orthogonality, or with $\gamma^i$, which will be contracted from both sides\footnote{Strictly speaking, this is true only for $\mathcal{H}$, since it has no free indices. For $S$ we may change the definition of the Lagrange multiplier $\bar{\psi}_t \rightarrow \bar{\psi}_t A$ to make it hold.} with $A(N)$ and all $A$s cancel due to $(A^{-1})_{IJ} A^{-1}\gamma^J A = \gamma_I$. Cancelling the $A$s makes $\mathcal{H}$ gauge invariant and $\mathcal{S}$ gauge covariant by inspection, if we replace all $\gamma^0$ by $i\slashed N$. Thus we are left with $\Theta$ and $\Lambda$, which are their own second class partners but Poisson commute with everything else, which can be seen as follows. For $\Theta$, note that $\mathcal{H}$, $\mathcal{S}$ and $\Lambda$ only depend on $\bar\eta^{JK}(A\rho_K + A^* \rho^*_K)$, which Poisson commutes with $\Theta$ due to the projector $\bar{\eta}$. For $\Lambda$, the situation again is more complicated. Remember that $\mathcal{H}$ and $\mathcal{S}$ in the time gauge only depended on $X^i \mathbb{P}_{ij} \rho^j$ for some $X^i$. Whatever $X^i$ may be, under (\ref{eq:OldVariablesNew}) it will be replaced by something of the form $A^{IJ} \bar{X}_J$ and the whole expression will become $\bar{X}_I A^{JI} \mathbb{P}_{JK} A^{KL} \bar{\eta}_{L}\m^{M} \left(A \rho_M + A^*\rho^*_M \right)$ with $\mathbb{P}_{IJ} = \eta_{IJ} - \frac{1}{D} \gamma_I \gamma_J$. Crucial for the following calculation is the property (\ref{eq:PropertiesOfA}), which will be used several times. Then we find that the generic term is Poisson commuting with $\Lambda$,
\ba
& &\left\{ \bar{X}_I A^{JI} \mathbb{P}_{JK} A^{KL} \bar{\eta}_{L}\m^M \left(A\rho_M + A^* \rho^*_M\right),   \gamma^{N} A_{NO} \bar{\eta}^{OP} (A \rho_P + A^*\rho^*_P) \right\}  \nonumber \\
&=&  -i\bar{X}_I A^{JI} \mathbb{P}_{JK} A^{KL} \bar{\eta}_{L}\m^M \left(\gamma^{N}\right)^T A_{NM} \nonumber \\
&=& -i\bar{X}_I A^{jI}  \mathbb{P}_{jk} A^{kL} \bar{\eta}_{L}\m^M \left(\gamma^{n}\right)^T A_{nM} \nonumber \\
&=& -i\bar{X}_I A^{jI}  \mathbb{P}_{jk} A^{kL} \gamma^{n} A_{nL}= -i \bar{X}_I A^{jI} \mathbb{P}_{jk} \gamma^{k} = 0\text{.}
\ea
The constraint algebra is summarised in table \ref{tab:Constraints}.
\begin{table}[h]
\renewcommand{\arraystretch}{2}\addtolength{\tabcolsep}{0.1pt}
\begin{center}
\begin{tabular}{|c|c|}
\hline 
First class constraints & Second class constraints \\
\hline 
$G^{IJ}$, $\tilde{\mathcal{H}}_a$, $\mathcal{H}$, $\mathcal{S}$, $S^a_{I\overline{M}}$ and $\mathcal{N}$ & $\Lambda$, $\Theta$, $\chi_I$ and $\chi$ \\
\hline 
\end{tabular}
\caption{List of first and second class constraints.}
\label{tab:Constraints}
\end{center}
\end{table}
\item By construction, $\mathcal{H}$ and $S$ reduce correctly if we choose time gauge $N^I = \delta^I_0$, which automatically implies $A_{IJ} \rightarrow (g_0)_{IJ} \in \text{SO}(D)_0$. Since the theory is SO$(D)_0$ invariant, a gauge transformation $g_0 \rightarrow 1$ can be performed, which implies $\rho^I = \rho^I_r$. From this one easily deduces that $G^{IJ}$ and $\tilde{\mathcal{H}}_a$ also reduce correctly. Since the theory was SO$(D+1)$ invariant in the beginning, these results do not depend on the gauge choice.
\item For the Dirac matrix, we find\footnote{Note that the Dirac matrix is block diagonal. Therefore, we do not need to consider the full Dirac matrix at once.}
\ba
 C_{IJ} &=&  \left\{ A \rho_I - (A \rho_I)^*,A \rho_J- (A \rho_J)^*\right\} \nonumber \\ &=& -i \eta_{IJ} \left[- AA^{\dagger} - A^*A^T \right] =  2i \mathbb{1} \eta_{IJ} \\
 (C^{-1})^{IJ} &=& -\frac{1}{i}\mathbb{1}\eta^{IJ} \\
 \left\{\rho_I,\rho_J\right\}_{DB} &=& - \left\{\rho_I,A \rho_K - (A \rho_K)^*\right\} (C^{-1})^{KL}  \left\{A \rho_L - (A \rho_L)^*,\rho_J\right\} = \nonumber \\
 &=& -\frac{i}{2} \eta_{IJ} A^{\dagger}A^* \text{,}
\ea
and for $\sigma$ analogously. We now can choose new variables which have simpler brackets. Motivated from the original replacement (\ref{eq:OldVariablesNew}), we define
\begin{alignat}{3}
  \rho_r^I &:= A^{IJ} A \rho_J\text{,}&~~~~ \sigma_r &:=A \sigma \text{,}\\
 \left(\rho_r^I\right)^* &=  A^{IJ} A^* \rho^*_J = A^{IJ} A^* ((A^*)^{-1}A\rho_J) = \rho_r^I \text{,}&~~~~~~~~ \sigma_r^* &= \sigma_r\text{,} ~~~~~~~~~~~~~~~~
\end{alignat}
with the Dirac brackets
\ba
\left\{\rho_r^I,\rho_r^J\right\}_{DB} &=&  \left\{A^{IK} A \rho_K,A^{JL} A \rho_L\right\}_{DB} = -\frac{i}{2} \eta^{IJ} AA^{\dagger}A^*A^T = -\frac{i}{2} \eta^{IJ}\mathbb{1}\text{,} \label{eq:DiracBracketRhoR} \\
\left\{\sigma_r,\sigma_r\right\} &=& i\frac{D}{2(D-1)}\mathbb{1}\text{.}
\ea
Thus, the Dirac brackets of the $\rho_r^I$, $\sigma_r$ are simple as are the reality conditions. Only the transformation behaviour of the new variables under SO$(D+1)$ rotations is complicated because of the appearance of the rotation $A$ in their definition. Note that also $\left\{P^I, P^J\right\}_{DB}$, $\left\{P^I, \rho^J_r \right\}_{DB}$ and $\left\{P^I, \sigma_r \right\}_{DB}$ will be non-zero. Therefore, we also choose a new variable $\tilde{P}^I$ with simple Dirac brackets, which can most easily be found by performing the symplectic reduction. After that, we can simply read it off the symplectic potential. We find using $\rho_I =  A_{JI} A^{-1} \rho^J_r$ and $\rho_I^{\dagger} = A_{JI} (\rho^{J}_r)^T A$ 
\ba
&\m& + i \rho_I^{\dagger}\dot{\rho}^I - i \frac{D-1}{D} \sigma^{\dagger} \dot{\sigma} + P^I \dot{N}_I \nonumber \\ 
&=& i A_{J}\m^I (\rho^{J}_r)^T A \dot{(A_{KI} A^{-1} \rho^K_r)} - i\frac{D-1}{D} \sigma^{T}_r A \dot{(A^{-1} \sigma_r)} + P^I \dot{N}_I  \nonumber \\
&=& i (\rho^{J}_r)^T \dot{\rho}_{Jr} -i \frac{D-1}{D} \sigma_r^T\dot{\sigma}_r +  P^I \dot{N}_I + \nonumber \\
&\m& + i \left( A_{J}\m^{L} (\rho^{J}_r)^T \frac{\partial A_{KL}}{\partial N_I} \rho^K_r +  (\rho^{J}_r)^T A \frac{\partial A^{-1}}{\partial N_I} \rho_{Jr} -\frac{D-1}{D} \sigma_r^T A \frac{\partial A^{-1}}{\partial N_I} \sigma_r \right) \dot{N}_I  \nonumber \\
&=&  i(\rho^{J}_r)^T \dot{\rho}_{Jr} -i \frac{D-1}{D} \sigma_r^T \dot\sigma_r + \tilde{P}^I \dot{N}_I \text{,}
\ea
with $\tilde{P}^I := P^I +  i A_{J}\m^L (\rho^{J}_r)^T \frac{\partial A_{KL}}{\partial N_I} \rho^K_r + i (\rho^{J}_r)^T A \frac{\partial A^{-1}}{\partial N_I} \rho_{Jr} - i \frac{D-1}{D} \sigma_r^T A \frac{\partial A^{-1}}{\partial N_I} \sigma_r $. It can be checked explicitly that $\tilde{P}^I$, expressed in the old variables $(P^I, N_J, \rho^{\dagger}_I, \rho^J, \sigma^{\dagger}, \sigma)$, Poisson commutes with the reality conditions and with itself, and therefore has nice Dirac brackets. For the spatial diffeomorphism constraint, a short calculation yields
\ba
\tilde{\mathcal{H}}_a &=& P^I \partial_a N_I - \frac{i}{2} \partial_a(\rho^{\dagger I}) \rho_I +\frac{i}{2} \rho^{\dagger I} \partial_a \rho_I +i\frac{D-1}{2D} \partial_a(\sigma^{\dagger}) \sigma -i\frac{D-1}{2D} \sigma^{\dagger} \partial_a \sigma + \hdots = \nonumber \\
&=& \tilde{P}^I \partial_a N_I + i(\rho^{I}_r)^T \partial_a \rho_{Ir} - i\frac{D-1}{D}\sigma^{T}_r \partial_a \sigma_r + \hdots \text{,}
\ea
which by inspection generates spatial diffeomorphisms on the new variables. The constraints $\Lambda$ and $\Theta$ become
\be
\Lambda = \gamma_i \rho_r^i \approx 0 ~~\text{and}~~ \Theta = \rho_r^0 \approx 0\text{,}
\ee
which look utterly non-covariant, but which by construction still Poisson commute with the SO$(D+1)$ Gau{\ss} constraint. It therefore has to have a complicated form. We find
\ba
G^{IJ} &=& 2P^{[I}N^{J]} + 2i\rho^{\dagger[I}\rho^{J]} + i\rho^{\dagger}_K [i\Sigma^{IJ}] \rho^K - i\left(\frac{D-1}{D}\sigma^{\dagger}\right) [i\Sigma^{IJ}] \sigma + \hdots  \nonumber \\
&=& 2\tilde{P}^{[I}N^{J]} + 2i \rho^{TK}_r A_{K}\m^{[I} A_{L}\m^{|J]} \rho^{L}_r + i \rho^{T}_{Kr} A[i\Sigma^{IJ}]A^{-1} \rho^K_r \nonumber  \\ 
& &- i\frac{D-1}{D}\sigma^{T}_r A [i\Sigma^{IJ}]A^{-1} \sigma_r  + 2i \left( A_{M}\m^{L} (\rho^{M}_r)^T \frac{\partial A_{KL}}{\partial N_{[I}} \rho^K_r + \right. \nonumber \\
& &+\left. (\rho^{N}_r)^T A \frac{\partial A^{-1}}{\partial N_{[I}} \rho_{Nr} - \frac{D-1}{D}\sigma_r^T A \frac{\partial A^{-1}}{\partial N_{[I}} \sigma_r  \right) N^{J]} + \hdots
\ea
\end{itemize}

Finally, we solve the remaining second class constraints $\Lambda$ and $\Theta$ which after a short calculations results in the final Dirac brackets
\be
\left\{\rho_r^i, \rho_r^j\right\}_{DB} = - \frac{i}{2} \mathbb{P}^{ij} \nonumber \text{,} ~~~
\left\{\rho_r^0, \rho_r^j\right\}_{DB} =0 \nonumber \text{,} ~~ ~
\left\{\rho_r^0, \rho_r^0\right\}_{DB}= 0 \nonumber \text{.} 
\ee
As a consistency check, we can consider
\be
\left\{\phi^i, \phi^j\right\}_{DB} = \left\{\rho_r^i + \frac{1}{D} \gamma^i \sigma_r,~ \rho_r^j+ \frac{1}{D} \gamma^j \sigma_r\right\}_{DB} = -( C^{-1})^{ij} \text{,}
\ee
which coincides with the Dirac brackets obtained in (\ref{eq:DiracBracket}).

The form of the Hamiltonian and supersymmetry constraints $\mathcal{H}$, $\mathcal{S}$ strongly depends on the Supergravity theory under consideration. Exemplarily, we cite the supersymmetry constraint in $D=3$, $N=1$ Supergravity from \cite{SawaguchiCanonicalFormalismOf} adapted to our notation,
\begin{eqnarray}
\mathcal{S} &=& - \frac{i}{2} \epsilon^{abc} \gamma_{5} \left[ \gamma_k e_a^k \hat{D}_b \left( \frac{1}{\sqrt[4]{q}} e_c^l \phi_l\right) + \hat{D}_b \left(  \frac{1}{\sqrt[4]{q}} \gamma_k e_a^k e_c^l \phi_l \right)  \right] \nonumber \\
&& + \frac{1}{2 \sqrt[4]{q}} \epsilon^{abc} \epsilon_{ijk} e_a^i \left(K'_b\ ^j + i \bar\phi_l \gamma^0 \gamma^{lj} E_{b}^m \phi_m\right) \gamma^k e_c^n \phi_n \text{,}
\end{eqnarray}
where $\hat{D}_a \phi_i = \partial_a \phi_i + \hat\omega_{aij} \phi^j + \frac{i}{2} \hat\omega_{akl} \Sigma^{kl} \phi_i$, $\hat\omega_{aij} = \Gamma_{aij} + \frac{i}{4 \sqrt{q}} e_a^k \left( \bar{\phi}_i \gamma_k \phi_j + 2 \bar{\phi}_{[i} \gamma_{j]} \phi_k \right)$, and $\Gamma_{aij}$ is the spin-connection annihilating the triad. An explicit expression for $\mathcal{S}$ in terms of the extended variables $(A,\pi,N,P,\rho,\rho^*, \sigma, \sigma^*)$ can be found using (\ref{eq:DecompositionFermions}), (\ref{eq:OldVariablesNew}). The corresponding constraint operator is obtained using the methods in section \ref{sec:Quantisation}, \ref{sec:KinematicalHilbertSpace} and \cite{ThiemannQSD5, BTTIII}.

\section{Background Independent Hilbert Space Representations for Majorana Fermions}
\label{sec:Quantisation}

Background independent Hilbert space representations for Dirac spinor fields were constructed 
in \cite{ThiemannKinematicalHilbertSpaces}. One may think that for the Rarita-Schwinger field 
or more generally for Majorana fermion fields one can reduce to this construction as follows:
Consider the following variables
\be
\xi^{I \alpha} = \frac{1}{\sqrt{2}} \left( \rho_r^{2\alpha+1} + i \rho_r^{2\alpha+2} \right), ~~~ \pi^{I \alpha} = \frac{-i}{\sqrt{2}} \left( \rho_r^{2\alpha+1} - i \rho_r^{2\alpha+2} \right), ~~~ \alpha = 1, \ldots, 2^{\lfloor (D+1)/2 \rfloor }  \text{,}
\ee
which have the non-vanishing Dirac brackets
\be
\left\{ \xi^{I \alpha}(x), \pi^{J \beta}(y) \right\} = -i \eta^{IJ} \delta^{\alpha \beta} \delta^{(D)}(x-y)
\ee
and the simple reality condition
\be
\bar{\pi} = - i \xi  \text{.}
\ee
The elements of the Hilbert space are field theoretic extensions of holomorphic (i.e. they only depend on $\theta^\alpha$) functions on the Gra{\ss}mann space spanned by the Gra{\ss}mann numbers $\theta^\alpha$ and their adjoints $\bar{\theta}^\alpha$, 
the operators corresponding to the phase space variables act as
\be
\hat{\xi} f := \theta f, ~~~ \hat{\pi} f := i \frac{d}{d \theta} f,
\ee
and the scalar product 
\be
<f, g> := \int e^{\bar{\theta} \theta} \bar{f} g \, d \bar{\theta} \, d \theta 
\ee
faithfully implements the reality conditions. There are, however, two drawbacks to this:\\
1. Due to the arbitrary split of the variables into two halves, the scalar product is not 
SO$(D)$ invariant which makes it difficult to solve the Gau{\ss} constraint.\\
2. The scalar product given above fails to implement the Dirac bracket resulting from 
the second class constraints, that is, $\{\rho^{r\alpha}_i,\rho^{r\beta}_j\}_{DB}=
-i/2\; \mathbb{P}^{\alpha\beta}_{ij}$. Recall that one must solve the second class constraints 
before quantisation, hence it is not sufficient to consider the quantisation of the Poisson bracket 
as was done above.\\
\\ 
In what follows we develop a background independent Hilbert space representation that 
is SO$(D)$ invariant, implements the Dirac bracket and is geared to real valued 
(Majorana) spinor fields. We begin quite generally with 
$N$ real valued Gra{\ss}mann variables $\theta_A,\;A=1,..,N;\; \theta_{(A} \theta_{B)}=0;\;
\theta_A^\ast=\theta_A$. We consider the finite dimensional, complex vector space $V$
of polynomials in the $\theta_A$ with complex valued coefficients. 
Notice that $f\in V$ depends on all real Gra{\ss}mann coordinates, it is not holomorphic
as in the case of the Dirac spinor field \cite{ThiemannKinematicalHilbertSpaces}.
Thus $\dim(V)=2^N$ is the complex dimension of $V$.
We may write a polynomial $f\in V$ in several equivalent ways which are useful in different contexts. Let $f^{(n)}_{A_1..A_n},\;0\le n\le N$ be a completely skew complex valued tensor 
(n-form) then $f$ can be written as 
\be \label{11}
f=\sum_{n=0}^N \;\frac{1}{n!}\; f^{(n)}_{A_1..A_n}\; \theta^{A_1}..\theta^{A_n} 
=\sum_{n=0}^N \; \sum_{1\le A_1<..<A_n\le N} \; f^{(n)}_{A_1..A_n}\; \theta^{A_1}..\theta^{A_n} \text{.}
\ee
An equivalent way of writing $f$ is by considering for $\sigma_k\in \{0,1\}$ 
and $A_1<..<A_n$ the relabelled coefficients
\be \label{22}
f_{\sigma_1..\sigma_N}:=f^{(n)}_{A_1..A_n},\;\;\; \sigma_k:=
\left\{ \begin{array}{cc}
1 & k\in\{A_1,..,A_n\} \\
0 & {\rm else} \text{.}
\end{array} 
\right.
\ee
It follows 
\be \label{33}
f=\sum_{\sigma_1,..,\sigma_N\in \{0,1\}}\; f_{\sigma_1..\sigma_N}\; \theta_1^{\sigma_1}..
\theta_N^{\sigma_N}
\ee
with the convention $\theta_A^0:=1$. 

On $V$ we define the obvious positive definite sesqui-linear form 
\be \label{44}
<f,f'>:=\sum^N_{n=0}\sum_{A_1<..<A_n} \; \overline{f^{(n)}_{A_1..A_N}}\; f^{(n)\prime}_{A_1.. A_N}
=\sum_{\sigma_1..\sigma_N}\; \overline{f_{\sigma_1..\sigma_n}} \; f'_{\sigma_1..\sigma_N}
\ee
as well as the operators 
\be \label{55}
[\theta_A \cdot f](\theta):=\theta_A\; f(\theta),\;\;\;\;
[\partial_A \cdot f](\theta):=\partial^l f(\theta)/\partial_A \text{,}
\ee
where the latter denotes the left derivative on Gra{\ss}mann space (see, e.g., 
\cite{HenneauxQuantizationOfGauge} for precise definitions). Notice the 
relations $\partial_{(A} \partial_{B)}=0,\;\; 2\partial_{(A} \theta_{B)}=\delta_{AB}$ which 
can be verified by applying them to arbitrary polynomials $f$. We claim that the 
operators (\ref{55}) satisfy the adjointness relation 
\be \label{5a}
\theta_A^\dagger=\partial_A \text{.}
\ee
The easiest way to verify this is to use the presentation (\ref{33}). We find explicitly
\ba \label{66}
\theta_A\cdot f  
&=& 
\sum_{\sigma_1,..,\sigma_N}\; f_{\sigma_1..\sigma_N}\;
(-1)^{\sigma_1+..+\sigma_{A-1}}\; \delta_{\sigma_A,0}\;\theta_1^{\sigma_1}..\theta_A..
\theta_N^{\sigma_N}
\nonumber\\
&=& 
\sum_{\sigma_1,..,\sigma_N}\; [f_{\sigma_1..\sigma_A-1..\sigma_N}\;
(-1)^{\sigma_1+..+\sigma_{A-1}}\; \delta_{\sigma_A,1}]\;\theta_1^{\sigma_1}..
\theta_N^{\sigma_N}
\nonumber\\
&=:& 
\sum_{\sigma_1,..,\sigma_N}\; \tilde{f}^A_{\sigma_1..\sigma_N}\;
\theta_1^{\sigma_1}..\theta_N^{\sigma_N} \text{,}
\nonumber\\
\partial_A\cdot f  
&=& 
\sum_{\sigma_1,..,\sigma_N}\; f_{\sigma_1..\sigma_N}\;
(-1)^{\sigma_1+..+\sigma_{A-1}}\; \delta_{\sigma_A,1}\;\theta_1^{\sigma_1}..\widehat{\theta_A}..
\theta_N^{\sigma_N}
\nonumber\\
&=& 
\sum_{\sigma_1,..,\sigma_N}\; [f_{\sigma_1..\sigma_A+1..\sigma_N}\;
(-1)^{\sigma_1+..+\sigma_{A-1}}\; \delta_{\sigma_A,0}]\;\theta_1^{\sigma_1}..
\theta_N^{\sigma_N}
\nonumber\\
&=:& 
\sum_{\sigma_1,..,\sigma_N}\; \hat{f}^A_{\sigma_1..\sigma_N}\;
\theta_1^{\sigma_1}..\theta_N^{\sigma_N} \text{,}
\ea
where the wide hat in the fourth line denotes omission of the variable. We conclude 
\ba \label{77}
<f,\theta_A f'> 
&=&  
\sum_{\sigma_1,..,\sigma_N}\; 
\overline{f_{\sigma_1..\sigma_N}}\;
\tilde{f}^{A\prime}_{\sigma_1..\sigma_N}
\nonumber\\
&=&  
\sum_{\sigma_1,..,\sigma_N}\; 
\overline{f_{\sigma_1..\sigma_N}\; (-1)^{\sigma_1+..+\sigma_{A-1}}}\;
\tilde{f}'_{\sigma_1..\sigma_A-1..\sigma_N} \delta_{\sigma_A,1}
\nonumber\\
&=&  
\sum_{\sigma_1,..,\sigma_N}\; 
\overline{f_{\sigma_1..\sigma_{A}+1..\sigma_N}\; (-1)^{\sigma_1+..+\sigma_{A-1}}
\delta_{\sigma_A,0}}\;
\tilde{f}'_{\sigma_1..\sigma_N}
\nonumber\\ 
&=& 
\sum_{\sigma_1,..,\sigma_N}\; 
\overline{\hat{f}^A_{\sigma_1..\sigma_N}}\;
\tilde{f}'_{\sigma_1..\sigma_N}=<\partial_A f,f'> \text{.}
\ea
Although not strictly necessary, it is interesting to see whether the scalar product 
(\ref{44}) can be expressed in terms of a Berezin integral, perhaps with a non trivial 
measure as in \cite{ThiemannKinematicalHilbertSpaces} for complex Gra{\ss}mann variables.
The answer turns out to be negative: The most general Ansatz for a ``measure'' is 
$d\mu=d\theta_1..d\theta_N, \; \mu(\theta)$ with $\mu\in V$ fails to reproduce (\ref{44}) 
if we apply the usual rule for the Berezin integral\footnote{Rather a linear functional on $V$
which is of course also a non Abelian Gra{\ss}mann algebra.}
$\int d\theta\; \theta^\sigma=\delta_{\sigma,1}$. Notice that from this we 
induce $\int d\theta_A\; d\theta_B=-\int d\theta_B \; d\theta_A$ as one quickly 
verifies when applying to $V$. However, there exists a non-trivial differential kernel
\be \label{88}
K:=(\theta_1+(-1)^{N-1}\partial_1)..(\theta_N+(-1)^{N-1}\partial_N)
\ee
such that 
\be \label{99}
<f,f'>=\int\; d\theta_N..d\theta_1\; f^\ast\;K\; f' \text{,}
\ee
where we emphasise that $f^\ast$ is the Gra{\ss}mann involution 
\be \label{1010}
f^\ast=\sum_{\sigma_1..\sigma_N} \overline{f_{\sigma_1..\sigma_N}}
\theta_N^{\sigma_N}..\theta_1^{\sigma_1}
=\sum_{\sigma_1..\sigma_N} \overline{f_{\sigma_1..\sigma_N}}
\;(-1)^{\sum_{k=1}^{N-1} \sigma_k \sum_{l=k+1}^N \sigma_l}\;
\theta_1^{\sigma_1}..\theta_N^{\sigma_N}
\ee
and not just complex conjugation of the coefficients of $f$. Notice also that due to total antisymmetry we may rewrite (\ref{99}) in the form 
\be \label{1111}
<f,f'>=\frac{(-1)^{N(N-1)/2}}{N!} \int d\theta_{A_1}..d\theta_{A_N} \;
f^\ast D_{A_1}..D_{A_N} \; f'
\ee
where 
\be \label{12}
D_A=\theta_A+(-1)^{N-1} \partial_A \text{.}
\ee 
The presentation (\ref{13}) establishes that the linear functional is invariant under 
U$(N)$ acting on $V$ by 
\be \label{13}
f\mapsto U\cdot f;\;\;\;\; [U\cdot f]^{(n)}_{A_1..A_N}= 
f^{(n)}_{B_1..B_N}
U_{B_1 A_1} .. U_{B_N A_N} \text{,}
\ee
which is of course also clear from (\ref{44}). Notice that (\ref{13}) formally corresponds to 
$\theta_A \mapsto U_{AB} \theta_B$ but this is not an action on real Gra{\ss}mann variables 
unless $U$ is real valued. If we want to have an action on the linear polynomials with real 
coefficients then we must restrict U$(N)$ to O$(N)$ or a subgroup thereof which will 
precisely the case in our application. In this case it is sufficient to restrict to real valued 
coefficients in $f$ and now the real dimension of $V$ is $2^N$.\\
\\ 
We sketch the proof that (\ref{88}) accomplishes (\ref{99}). We introduce the notation
for $k=1,..,N$
\be \label{14}
F_{\sigma_1 .. \sigma_k}:=\sum_{\sigma_{k+1}.. \sigma_N} \; f_{\sigma_1..\sigma_N}\;
\theta_{k+1}^{\sigma_{k+1}}.. \theta_N^{\sigma_N} \text{,}
\ee
whence $F_{\sigma_1..\sigma_N}=f_{\sigma_1..\sigma_N}$. 
Notice that $F_{\sigma_1..\sigma_k}$ no longer depends on $\theta_1,..,\theta_k$.
Using this we compute 
with $d^N\theta:=d\theta_N..d\theta_1$ and using anticommutativity at various places
\ba \label{15} 
&& \int \; d^N\theta\;f^\ast\; K\; f'
\nonumber\\
&=& 
\int \; d^N\theta\; [F_0^\ast+F_1^\ast \; \theta_1]\;
(\theta_1+(-1)^{N-1}\partial_1)\;D_2\;..\; D_N\; [F'_0+\theta_1 \; F'_1]
 \nonumber\\
 &=& 
\int \; d^N\theta\; \Big\{F_0^\ast\; (-1)^{N-1}\; D_2..D_N (\theta_1+(-1)^{N-1}\partial_1)
[F'_0+\theta_1 F'_1]
\nonumber\\
&&
~~~~~~~~~~~~+F_1^\ast \;(-1)^{N-1}\;D_2\;..\; D_N\;\theta_1\partial_1 [F'_0+\theta_1 \; F'_1] \Big\}
 \nonumber\\ 
 &=& 
\int \; d^N\theta\; \Big\{F_0^\ast\; (-1)^{N-1}\; D_2..D_N \;
[\theta_1 F'_0+(-1)^{N-1} F'_1]
+F_1^\ast \;(-1)^{N-1}\;D_2\;..\; D_N\;\theta_1\; F'_1\Big\}
 \nonumber\\  
 &=& 
\int \; d^N\theta\; \Big\{F_0^\ast\; \theta_1\; D_2..D_N \;F'_0
+F_1^\ast \; \theta_1 \;D_2\;..\; D_N \; F'_1\Big\} \text{,}
 \ea
where we used that the second term no longer is linear in $\theta_1$ 
 and therefore drops out from the Berezin integral. The calculation explains why the 
 factor $(-1)^{N-1}$ in (\ref{12}) is necessary. 
 
 Next consider the first term in the last line of (\ref{15}). We have 
 \ba \label{16}
&&  \int \; d^N\theta\; F_0^\ast\; \theta_1\; D_2..D_N \;F'_0  
\nonumber\\
&=& 
\int \; d^N\theta\; [F_{00}^\ast+F_{01}^\ast \theta_2] \; \theta_1\; 
(\theta_2+(-1)^{N-1}\partial_2)\; D_3..D_N \;
[F'_{00}+\theta_2 F'_{01}]
\nonumber\\
&=& 
\int \; d^N\theta\; \Big\{F_{00}^\ast\;(-1)^{N-2}\;\theta_1\; D_3..D_N\;(\theta_2+(-1)^{N-1}\partial_2)
[F'_{00}+\theta_2 F'_{01}]
\nonumber\\
&&
~~~~~~~~~~~~+F_{01}^\ast  \; (-1)^{N-2}\; \theta_1\; D_3..D_N \;\theta_2\partial_2\;
[F'_{00}+\theta_2 F'_{01}] \Big\}
\nonumber\\
&=& 
\int \; d^N\theta\; \Big\{F_{00}^\ast\;(-1)^{N-2}\;\theta_1\; D_3..D_N\;
[\theta_2 F'_{00}+(-1)^{N-1} F'_{01}]
+F_{01}^\ast  \; (-1)^{N-2}\; \theta_1\; D_3..D_N \;\theta_2\; F'_{01}] \Big\}
\nonumber\\ 
&=& 
\int \; d^N\theta\; \Big\{F_{00}^\ast \; \theta_1\theta_2\; D_3..D_N\; F'_{00}
+F_{01}^\ast  \; \theta_1 \theta_2\; D_3..D_N \; F'_{01}\Big\} \text{.}
\ea 
 Similarly for the second term in (\ref{15})
\be \label{17}
 \int \; d^N\theta\; F_1^\ast\; \theta_1\; D_2..D_N \;F'_1  
 =
\int \; d^N\theta\; \Big\{F_{10}^\ast \; \theta_1\theta_2\; D_3..D_N\; F'_{10}
+F_{11}^\ast  \; \theta_1 \theta_2\; D_3..D_N \; F'_{11}\Big\} \text{.} 
\ee
It is transparent how the computation continues: We continue expanding 
$F_{\sigma_1..\sigma_k}=F_{\sigma_1..\sigma_k 0}+\theta_{k+1} 
F_{\sigma_1 .. \sigma_k 1}$ and see by exactly the same computation as above 
that\footnote{A strict proof would proceed by induction which we leave as an easy 
exercise for the interested reader.} the signs match up to the effect that
\be \label{18}
 \int \; d^N\theta\; F_{\sigma_1..\sigma_k}^\ast\; \theta_1..\theta_k\; D_{k+1}..D_N \;
 F'_{\sigma_1..\sigma_k}  
 =\sum_{\sigma_{k+1}}\;
 \int \; d^N\theta\; F_{\sigma_1..\sigma_{k+1}}^\ast\; \theta_1..\theta_{k+1}\; D_{k+2}..D_N \;
 F'_{\sigma_1..\sigma_{k+1}}  \text{,}
\ee 
from which the claim follows using $\int\; d^N\theta \; \theta_1..\theta_N=1$.
 In our applications $N$ will be even so that 
$D_A=\theta_A-\partial_A$. \\
\\
For our application to the Rarita-Schwinger field we consider the compound index
$A=(j,\alpha),\;j=1,..,D;\;\alpha=1,..,2^{[(D+1)/2\rfloor}$ or just $A=\alpha$ 
whence $N=DM$ or $N=M$ is even. We consider
the auxiliary operator 
\be \label{19}   
\hat{\rho}_j^\alpha:=\frac{\sqrt{\hbar}}{2}[\theta_j^\alpha+\partial_j^\alpha] \text{,}
\ee
which by virtue of  (\ref{5a}) is self adjoint and satisfies the anticommutator relation
\be \label{20}
[\hat{\rho}_j^\alpha,\hat{\rho}_k^\beta]_+=\frac{\hbar}{2} \delta_{jk}\;\delta^{\alpha\beta} \text{.}
 \ee
 However, $\hat{\rho}_j^\alpha$ is not yet a representation of $\rho^{r\alpha}_j$ which 
 satisfies the Dirac antibracket $\{\rho_j^{r\alpha},\rho_k^{r\beta}\}_{{\rm DB}}=
 -\frac{i}{2}\mathbb{P}^{\alpha\beta}_{jk}$ and the 
 reality condition $(\rho_j^{r\alpha})^\ast=\rho_j^{r\alpha}$.  Similarly, 
 $\{\sigma_\alpha,\sigma_\beta\}_{{\rm DB}}=i\frac{D}{2(D-1)}\delta_{\alpha\beta},\;\;\sigma_\alpha^\ast=
 \sigma_\alpha$. Correspondingly, what we need is a representation $\pi(\rho_j^\alpha),\;\pi(\sigma_\alpha)$
 of the abstract CAR $^\ast$-algebra  defined by canonical quantisation, that is, 
 \be \label{2222}
 [\rho_j^{r\alpha},\rho_k^{r\beta}]_+=\frac{\hbar}{2} \mathbb{P}^{\alpha\beta}_{jk},\;\;\;\;
 (\rho_j^{r\alpha})^\ast=
 \rho_j^{r\alpha},\;\;\;\;
 [\sigma^r_\alpha,\sigma^r_\beta]_+=\frac{D \hbar}{2(D-1)} \delta_{\alpha\beta},\;\;\;\;
 (\sigma^r_\alpha)^\ast
 =\sigma^r_\alpha
 \ee 
 all other anticommutators vanishing\footnote{This corresponds to the quantisation rule
 that the anticommutator is $+i\hbar$ times the Dirac bracket in the $\rho$ sector 
 and $-i\hbar$ times the Dirac bracket in the $\sigma$ sector. This is the only possible 
 choice of signs because the anticommutator of the same operator which in our case 
 is self adjoint is a positive operator. The other choice of signs would yield a 
 mathematical contradiction.}.
 Fortunately, using that $P_{jk}^{\alpha\beta}$ is a real valued 
 projector (in particular symmetric and positive semidefinite) we can now write the following 
 faithful representation
 of our abstract $^\ast$-algebra (\ref{2222}) on the Hilbert ${\cal H}=V_{DM}\otimes V_M$ defined above:
 \ba \label{23}
 \pi(\rho_j^{r\alpha}):=\mathbb{P}_{jk}^{\alpha\beta} \hat{\rho}_k^\beta,\;\;\;\;
 \pi(\sigma^r_\alpha):=\frac{1}{2} \sqrt{\frac{D \hbar}{D-1}}[\theta_\alpha+\partial_\alpha] \text{.}
 \ea

 ~\\
 So far we have considered just one point on the spatial slice corresponding to a quantum
 mechanical system. The field theoretical generalisation now proceeds exactly as in 
 \cite{ThiemannKinematicalHilbertSpaces} and consists in considering copies ${\cal H}_x$
 of the Hilbert space just constructed, one for every spatial point $x$ and taking 
 as representation space either the inductive limit of the finite tensor products 
 ${\cal H}_{x_1,..,x_n}=
 \otimes_{k=1}^n\; {\cal H}_{x_k}$ \cite{ThiemannModernCanonicalQuantum} or the infinite tensor product
 \cite{ThiemannGCS4} ${\cal H}=\otimes_x {\cal H}_x$ of which the former is just a tiny subspace.
The $^\ast$-algebra  (\ref{2222}) is then simply extended by adding labels $x$ to the 
operators and to ask that anticommutators between operators at points $x,y$ be proportional to
$\delta_{x,y}$ in agreement with the classical bracket. It is easy to see that adding 
the label $x$ to (\ref{23}) correctly reproduces this Kronecker symbol and that 
they satisfy all relations on the Hilbert space\footnote{In the case of the inductive limit,
a vector in $v\in {\cal H}_{x_1,..,x_n}$ is embedded in 
any larger ${\cal H}_{x_1,.., x_n,y_1,..,y_m}$ by $v\mapsto v\otimes \;\otimes^m 1$ where 
$1$ is the constant polynomial equal to one. This way any operator at $x$ acts 
in a well defined way on any vector in the Hilbert space.}. Finally notice that 
the corresponding scalar product is locally SO$(D)$ invariant.

\section{Generalisations to Different Multiplets}

\subsection{Majorana Spin $1/2$ Fermions}

The above construction generalises immediately to Majorana spin $1/2$
fermions which are also present in Supergravity theories, e.g.
$D+1=9+1$, $N=2a$ non-chiral Supergravity \cite{GianiN=2SupergravityIn}. They are described by actions
of the type
\be
S_{\text{Majorana, } 1/2} = \int d^{D+1}X \, i \bar{\lambda} \gamma^\mu 
D_\mu \lambda
\ee
which, using time gauge and a real representation for the
$\gamma$-matrices, lead to the canonical brackets $\{ \lambda_\alpha,
\lambda_\beta \} \sim i \delta_{\alpha \beta}$ with the reality conditions
$\lambda^* = \lambda$. They can thus also be treated with the above
techniques by substituting $\rho_i$ with $\lambda$ and removing the
$A_{IJ}$ matrices as well as the $\bar{\eta}^{IJ}$ projectors.

\subsection{Mostly Plus / Mostly Minus Conventions}

The convention used for the internal signature, i.e. mostly plus or
mostly minus and the associated purely real or purely imaginary
representations of the $\gamma$-matrices, does not interfere with the
above construction. The important property we are using is the reality
of $i \Sigma_{IJ}$ for SO$(1,D)$, i.e. that the Gau{\ss} constraint is
consistent with real spinors. The substitution $\gamma_I \rightarrow i
\gamma_I$ necessary when changing the signature convention does not
influence these considerations.

\subsection{Weyl Fermions}
In dimensions $D+1$ even, we also need to consider the case of Weyl
fermions. To this end, we define
\be
\gamma_{\text{five}} := i^{\frac{D(D+1)}{2} + 1} \gamma^L_0 \gamma_1
\hdots \gamma_{D}
\ee
with the properties $\gamma_{\text{five}}^2 = 1$,
$\gamma_{\text{five}}^{\dagger} = \gamma_{\text{five}}$ and
$\left[\gamma_I, \gamma_{\text{five}}\right]_+ = 0$ (which follows from
our conventions for the gamma matrices $(\gamma_0^L)^2 = -1$,
$\gamma_i^2 = 1$, $\gamma_I^{\dagger} = \eta_{II} \gamma_I$). We
introduce the chiral projectors
\be
\mathcal{P}^{\pm} = \frac{1}{2} \left( 1 \pm \gamma_{\text{five}} \right) \text{,}
\ee
which fulfil the relations $\mathcal{P}^{\pm}\mathcal{P}^{\pm} =
\mathcal{P}^{\pm}$, $\mathcal{P}^{\pm} \mathcal{P}^{\mp} = 0$,
$\mathcal{P}^{+}+ \mathcal{P}^{-} = 1$ and
$(\mathcal{P}^{\pm})^{\dagger} = \mathcal{P}^{\pm}$. These follow
directly from the properties of $\gamma_{\text{five}}$.

\subsubsection{Spin $1/2$ Dirac-Weyl Fermions}
The kinetic term of the action for a chiral Dirac spinor is given by
\be S_{\text{F}} = -\int_\mathcal{M} d^{D+1} X \left(\frac{i}{2}
\overline{\Psi} e^\mu_I \gamma^I D_\mu \mathcal{P}^+ \Psi - \frac{i}{2}
\overline{ D_\mu \Psi} e^\mu_I \gamma^I \mathcal{P}^+ \Psi \right)
\text{.} \label{eq:action} \ee
The $3+1$ split is performed analogous to \cite{BTTIII}.
Choosing time gauge, we obtain the non-vanishing Poisson brackets
\be
\left\{ \Psi^{\pm}_{\alpha}, \Pi^{\pm}_{\beta} \right\} =
-\mathcal{P}^{\pm}_{\alpha \beta}\text{,}
\ee
where $\Pi^{\pm}_{\beta} = -i (\Psi^{\pm})^{\dagger}_{\beta}$, and the
first class constraint
\be
\chi_{\alpha} := \Pi^{-}_{\alpha} \text{,}
\ee
where we used the notation $\Psi^{\pm} := \mathcal{P}^{\pm} \Psi$. The
first class property of this constraint follows from the fact that the
action (\ref{eq:action}) and therefore all resulting constraints do not
depend on $\Psi^{-}$ at all. In the quantum theory, the Hilbert space
for the chiral fermions can be constructed similar to the case of
non-chiral ones \cite{ThiemannKinematicalHilbertSpaces}. We obtain a
faithful representation of the Poisson algebra by replacing the
operators $\hat \theta_{\alpha}$ (acting by multiplication) and $\hat
{\bar{ \theta}}_{\alpha} = \frac{d}{d\theta_{\alpha}}$ defined in
\cite{ThiemannKinematicalHilbertSpaces} by $\hat \theta^+_{\alpha} :=
\mathcal{P}^+_{\alpha \beta} \hat{\theta}_{\beta}$ and $\hat{\bar{
\theta}}^+_{\alpha} := \hat {\bar{ \theta}}_{\beta} \mathcal{P}^+_{
\beta \alpha}$, as can be seen by
\be
\left[\hat \theta^+_{\alpha}, \hat{\bar{ \theta}}^+_{\beta}\right]_+ =
\mathcal{P}^+_{\alpha \beta}\text{.}
\ee
The reality conditions are implemented if we use the unique measure
constructed in \cite{ThiemannKinematicalHilbertSpaces}. We then have to
impose the condition
\be
\hat{\bar{\theta}}^-_{\alpha} f(\{\theta_{\beta}\}) = 0 \text{,}
\ee
which restricts the Hilbert space to functions $f$ such that
$f(\{\theta_{\alpha}\})= f(\{\mathcal{P}^{+}_{\alpha \beta}
\theta_{\beta}\})$. Classically, observables do not depend on
$\Psi^-_{\alpha}$. In the quantum theory, they become operators which do
not contain $\hat{\theta}^-_{\alpha}$ and therefore commute with
$\hat{\bar{\theta}}^-_{\alpha}$.

\subsubsection{Spin $3/2$ Majorana-Weyl Fermions}
Majorana-Weyl spin $3/2$ fermions appear in chiral Supergravity
theories, e.g., $D+1=9+1$, $N=1$ \cite{ChamseddineN=4}. In general, in a
real representation ($\gamma_I^T = \eta_{II} \gamma_I$) or in a
completely imaginary representation ($\gamma_I^T = -\eta_{II} \gamma_I$)
we have $\gamma_{\text{five}}^T = (-1)^{\frac{D(D+1)}{2}+1}
\gamma_{\text{five}}$. Therefore, if $\frac{D+1}{2}$ is odd, we have
$(\mathcal{P}^{\pm})^T = \mathcal{P}^{\pm}$, and if $\frac{D+1}{2}$ is
even, $(\mathcal{P}^{\pm})^T = \mathcal{P}^{\mp}$. In the case at hand
($D=9$), there exists a real representation and the chiral projectors
will be symmetric, $(\mathcal{P}^{\pm})^T = \mathcal{P}^{\pm}$. Again,
we will just consider the kinetic term for a chiral Rarita-Schwinger field,
\be
S = \int_{\mathcal{M}} d^{D+1}X \left(is ~ e \overline{\psi}_{\mu}
\gamma^{\mu \rho \sigma} D_{\rho} \mathcal{P}^{+}\psi_{\sigma} \right)
\text{.}
\ee
The $3+1$ split is performed like above. We find the second class
constraint $\pi_i^{+} = i (\phi_j^+)^T \gamma^{ji} \mathcal{P}^+$ and the
first class constraint $\pi^{-}_i = 0$. We introduce a second class
partner $\phi_i^- = 0$ for the first class constraint. Then we can solve
all the constraints using the Dirac bracket
\be
\left\{\phi_i^+, \phi_j^+\right\}_{DB} = - \mathcal{P}^+ (C^{-1})_{ij}
\mathcal{P}^+ \text{,} \label{eq:DiracBracketsMajorana}
\ee
and all other brackets are vanishing. From here, we can copy the
enlargement of the internal space from above, which results in the same
theory with all variables projected with $\mathcal{P}^+$. (Note that
equations like e.g. $\rho_i^+ = \frac{1}{2} A_{iJ} \bar{\eta}^{JK}
\mathcal{P}^+ \left(A\rho_K + A^* \rho^*_K\right) = \frac{1}{2} A_{iJ}
\bar{\eta}^{JK} \left(A\rho^+_K + A^* (\rho^+)^*_K\right)$ are
consistent. This can be seen by the fact that the matrix $A(N)$ can be
written as an infinite sum of even powers of gamma matrices, $A(N)
\propto \exp(i {\Lambda_{IJ}(N) \Sigma^{IJ}})$ and therefore it commutes
with the projectors $\mathcal{P}^{\pm}$.) The quantisation of the
resulting theory with variables $\rho^+_r\m_I$ and $\sigma^+$ is similar
to the non-chiral case, with chiral projectors $\mathcal{P}^+$ added in
observables, and modifications of the Hilbert space similar to the ones
given above for Dirac-Weyl fermions.

\section{Conclusions}

In the present paper we have demonstrated that the complications arising when trying 
to extend canonical Supergravity in the time gauge from the gauge group SO$(D)$ to 
SO$(D+1)$ in order to achieve a seamless match to the canonical connection formulation
of the graviton sector outlined in \cite{BTTI,BTTII} can be resolved. Since we worked with a Majorana representation of the $\gamma$-matrices, our analysis is restricted to those dimensions where this representation is available, which, however, covers many interesting supergravity theories ($d=4, 8,9,10,11$). The price to pay for the enlargement of the gauge group is that
the phase space requires an additional normal field $N$ and that 
the constraints depend non trivially on a matrix $A(N)$ which transforms in a complicated fashion
under SO$(D+1)$ but which in the present formulation is crucial in order to formulate 
the reality conditions for the Majorana fermions in the SO$(D+1)$ theory.

One would expect that the field $N$ is superfluous and that the matrix $A(N)$ would simply drop out when performing
an extension to SO$(1,D)$ because then no non trivial reality conditions need to be imposed.
One would expect that one only needs the quadratic and not the linear simplicity constraint and 
that, just as it happened in the graviton sector \cite{BTTI,BTTII}, the Hamiltonian phase space extension
method simply coincides with the direct Hamiltonian formulation obtained by an 
$D+1$ split of the SO$(1,D)$ action followed by a gauge unfixing step in order to obtain a first 
class formulation. Surprisingly, this is not the case. The basic difficulty is that when 
performing the $D+1$ split without time gauge, the symplectic structure turns 
out to be unmanageable. A treatment similar to the one carried out in this paper is possible
but turns out to be of similar complexity. It therefore appears that there is no advantage 
of the SO$(1,D)$ extension as compared to the SO$(1+D)$ even as far as the classical 
theory is concerned. We hope to communicate our findings in a forthcoming publication. 
Of course, the quantum theory of the SO$(1,D)$ extension is beyond 
any control at this point.

The solution to the tension presented in this paper, between having real Majorana spinors coming from SO$(1,D)$ 
on the one hand and 
an SO$(D+1)$ extension of the theory which actually needs complex valued spinors on the other,
is most probably
far from unique nor the most elegant one. Several other solutions have suggested themselves 
in the course of our analysis but the corresponding reformulation is not yet complete
at this point. Hence, we may revisit this issue in the future and simplify the presentation. Furthermore, it would be interesting to investigate if the extensions of the gauge group $SO(D) \rightarrow SO(D+1)$ also is possible in the case of symplectic Majorana fermions, which would permit access to even more supergravity theories.      

To the best of our knowledge, the background independent Hilbert space representation of the Rarita-Schwinger 
field presented in section 4 is also new. Apart from the fact that this has to be done 
for half-density valued Majorana spinors whose tensor index is transformed into an external
one by contracting with a vielbein, as compared to Dirac spinors there is no 
representation in terms of holomorphic functions \cite{ThiemannKinematicalHilbertSpaces}
 of the Gra{\ss}mann variables and one had to deal with the non trivial Dirac bracket.\\
\\
\\
\\
{\bf\large Acknowledgements}\\
NB and AT thank Christian Fitzner for discussions about fermionic variables and the German National Merit Foundation for financial support. We thank two anonymous referees for helpful comments. The part of the research performed at the Perimeter Institute for Theoretical Physics was supported in part by funds from the Government of Canada through NSERC and from the Province of Ontario through MEDT.

\newpage

\begin{appendix}

\section{Linear Simplicity Constraints}

As outlined in the main text, the most convenient SO$(D+1)$ extension of SO$(D)$ Lorentzian
Supergravity in the time gauge employs a normal vector field $N$ for which we have to provide 
symplectic structure, additional constraints and its interplay with the quadratic simplicity constraint in order to make sure that the physical content of the theory remains unaltered.
In effect, the results of \cite{BTTI, BTTII} are reformulated in terms of a linear simplicity constraint in the spirit of the new Spin Foam models \cite{EngleTheLoopQuantum, LivineNewSpinfoamVertex, EngleFlippedSpinfoamVertex, EngleLoopQuantumGravity, FreidelANewSpin, KaminskiSpinFoamsFor}. Therefore, a dynamical unit-length scalar field $N^I$ will be introduced, which - if the simplicity constraints hold - has the interpretation of the normal to the (spatial pullback of the spacetime $(D+1)$-) vielbein in the internal (Lorentzian or Euclidean) space. It will be shown that the constraints comprise a first class system and that the theory in any dimension is equivalent to the ADM formulation of General Relativity. The results are shown to extend to coupling of fermionic matter treated in \cite{BTTIV}. Like in \cite{BTTI,BTTII}, we can choose either SO$(D+1)$ or SO$(1,D)$ as gauge group for Lorentzian gravity.
However,  only for the compact case, we are able to construct the Hilbert space $\mathcal{H}_N$ for the normal field $N^I$.
In a companion paper \cite{BTTV},  we will comment on the implementation of the linear simplicity constraint operators on the Hilbert space $\mathcal{H}_T = \mathcal{H}_\text{grav} \otimes \mathcal{H}_N$.

\subsection{Introductory Remarks}

In \cite{BTTI,BTTII}, gravity in any dimension $D$ has been formulated as a gauge theory of SO$(1,D)$ or of the compact group SO$(D+1)$, irrespective of the space time signature. The resulting theory has been obtained by two different routes, a Hamiltonian analysis of the Palatini action making use of the procedure of gauge unfixing\footnote{See \cite{MitraGaugeInvariantReformulation, AnishettyGaugeInvarianceIn, VytheeswaranGaugeUnfixingIn} for original literature on gauge unfixing.}, and on the canonical side by an extension of the ADM phase space. The additional constraints appearing in this formulation, the simplicity constraints, are well known. They constrain bivectors to be simple, i.e. the antisymmetrised product of two vectors. Originally introduced in Plebanski's \cite{PlebanskiOnTheSeparation} formulation of General Relativity as constrained $BF$ theory in $3+1$ dimensions, they have been generalised to arbitrary dimension in \cite{FreidelBFDescriptionOf}. Moreover, discrete versions of the simplicity constraints are a standard ingredient of the covariant approaches to Loop Quantum Gravity called Spin Foam models \cite{BarrettRelativisticSpinNetworks, EngleFlippedSpinfoamVertex, FreidelANewSpin} and recently were also used in group field theory \cite{BaratinGroupFieldTheory}. Two different versions of simplicity constraints are considered in the literature, which are either quadratic or linear in the bivector fields. The quantum operators corresponding to the quadratic simplicity constraints have been found to be anomalous both in the covariant \cite{EngleLoopQuantumGravity} as well as in the canonical picture \cite{WielandComplexAshtekarVariables, BTTIII}. On the covariant side, this lead to one of the major points of critique about the Barrett-Crane model \cite{BarrettRelativisticSpinNetworks}: The anomalous constraints are imposed strongly\footnote{Strongly here means that the constraint operator annihilates physical states, $\hat C \left|\psi\right\rangle = 0 ~ \forall \left| \psi \right\rangle \in \mathcal{H}_{\text{phys}}$.}, which may imply erroneous elimination of physical degrees of freedom \cite{DiracLecturesOnQuantum}. This lead to the development of the new Spin Foam models \cite{EngleTheLoopQuantum, LivineNewSpinfoamVertex, EngleFlippedSpinfoamVertex, EngleLoopQuantumGravity, FreidelANewSpin, KaminskiSpinFoamsFor}, in which the quadratic simplicity constraints are replaced by linear simplicity constraints. The linear version of the constraint is slightly stronger than the quadratic constraint, since in $3+1$ dimensions the topological solution is absent. The corresponding quantum operators are still anomalous (unless the Immirzi parameter takes the values $\beta = \pm \sqrt{\zeta}$, where $\zeta$ denotes the internal signature). Therefore, in the new models (parts of) the simplicity constraints are implemented weakly to account for the anomaly.

To make contact to the covariant formulation, it is therefore of interest to ask whether, from the canonical point of view, $(a)$ the theory of \cite{BTTI,BTTII} can be reformulated using a linear simplicity constraint, and $(b)$ if so, whether the linear version of the constraint can be quantised without anomalies. Both of these questions will be answered affirmatively, the answer to $(a)$ in this appendix and the answer to $(b)$ in our companion paper \cite{BTTV}. As we have shown in the present paper, the use of the linear simplicity constraints (already at the classical level) 
is probably the most convenient approach towards constructing a connection formulation for Supergravity theories in $D+1$ dimensions with compact gauge group.
To answer $(a)$ we will follow the second route as in \cite{BTTI} and construct the theory with linear simplicity constraint by an extension of the ADM phase space.

Note that the linear constraints already have been introduced in a continuum theory in \cite{GielenClassicalGeneralRelativity}, yet the considerations there are rather different. The authors reformulate the action of the Plebanski formulation of General Relativity using constraints which involve an additional three form and which are linear in the bivectors, without giving a Hamiltonian formulation. This paper on the other hand will deal exclusively with the Hamiltonian framework.

Notice that we denote by ($s$) the space time signature and by ($\zeta$) the internal signature, which can be chosen independently as in \cite{BTTI}. In particular, the gauge group SO$(\eta)$ (with $\eta = \text{diag}(\zeta,1,1,...)$) can be chosen compact, irrespective of the space time signature.
This will be exploited when quantising the theory in \cite{BTTV}, where we fix $\zeta = 1$ and therefore do not have to bother with the non-compact gauge group SO$(1,D)$. There, we employ the
 Hilbert space representation for the normal field derived in section \ref{sec:KinematicalHilbertSpace} of this paper  and then we find quantum operators corresponding to the linear simplicity constraint and show that these operators $(b)$ actually are {\it{of the first class}} and therefore can be implemented strongly.
\\
\\
This appendix is organised as follows. Since the construction of the new theory follows neatly the treatment in \cite{BTTI}, in section \ref{sec:QuadraticSimplicity} we will shortly review the extensions of the ADM phase space introduced there, highlighting those details which will become important in the case of linear simplicity constraints. In section \ref{sec:LinearSimplicity} the new theory is presented and proved to be equivalent to the ADM formulation, which already implies that solving the linear simplicity constraints classically (section \ref{sec:Solution}) leads back to the (extended) ADM phase space and its constraints. Next, we show that the framework can be extended to coupling of fermionic matter (section \ref{sec:Fermions}). Finally, we construct a background 
independent Hilbert space representation for the normal field $N^I$ in section
(\ref{sec:KinematicalHilbertSpace}) which exploits the fact that $N^I$ on-shell is a unit vector and therefore valued in a compact set.

\subsection{Review: Quadratic Simplicity Constraints}
\label{sec:QuadraticSimplicity}

\subsubsection{Step 1: $\left\{ K_{aIJ}, \pi^{bKL}\right\}$ - Theory}
\label{sec:KPi}
In \cite{BTTI}, the ADM phase space is extended using the variables $\pi^{aIJ}$ and $K_{bKL}$, which are related to the ADM variables via
\ba \pi^{aIJ} \pi^{b}\m_{IJ} &:=& 2 \zeta q q^{ab}  \label{eq:metric} \text{,} \\
K^{ab} &:=& - \frac{s}{4 \sqrt{q}} \pi^{bKL} K_{cKL} q^{ac}\left(\pi\right) \text{,} \\
P^{ab} &=& -s \sqrt{q} \left( K^{ab} - q^{ab} K^c \m_c\right) = \frac{1}{4} \left( q^{ac}\left(\pi\right) \pi^{bKL} K_{cKL} - q^{ab}\left(\pi\right) \pi^{cKL} K_{cKL} \right) \text{.} \ea
The ADM constraints expressed in this variables\footnote{Note that by calculating the determinant of equation (\ref{eq:metric}), we can express both $q$ and $q^{ab}$ in terms of $\pi^{aIJ}$, and via the formula for the inverse matrix we also obtain an expression for $q_{ab}(\pi)$. All metric-related quantities, like e.g. the Levi-Civita connection $\Gamma_{ab}^c := \frac{1}{2} q^{cd} \left( \partial_a q_{bd} + \partial_b q_{ad} - \partial_d q_{ab} \right)$ and the Ricci scalar $R$, can now be expressed in terms $\pi^{aIJ}$ and will automatically be understood as functions of $\pi^{aIJ}$ in the following. To keep notation simple, the $\pi^{aIJ}$ - dependence will not be made explicit.} are given by
\begin{eqnarray}
 \mathcal{H}_a &=& -2 q_{ac} \nabla_b P^{bc} = -\frac{1}{2} \nabla_b \left( K_{aIJ} \pi^{bIJ} - \delta_a^b K_{cIJ} \pi^{cIJ} \right) \text{,} \\
 \mathcal{H} &=& - \left[ \frac{s}{\sqrt{q}} \left( q_{ac} q_{bd} - \frac{1}{D-1} q_{ab} q_{cd} \right) P^{ab} P^{cd} + \sqrt{q} R \right]  \nonumber \\
 						&=& - \frac{s}{8\sqrt{q}} \left( \pi^{[a|IJ} \pi^{b]KL} K_{bIJ} K_{aKL}\right) - \sqrt{q} R \text{,}
\end{eqnarray}
where $\nabla_a$ is the covariant derivative annihilating the spatial metric. In order to have the right number of physical degrees of freedom, the Gau{\ss} and the (quadratic) simplicity constraints are introduced
\ba
G^{IJ} &:=& 2 K_{a}\m^{[I}\m_{K} \pi^{aK|J]}\text{,} \label{eq:gauss}\\
S^{ab}_{\overline{M}} &:=& \frac{1}{4} \epsilon_{IJKL\overline{M}} \pi^{aIJ} \pi^{bKL} \text{.} \label{eq:simplicity} 
\ea
Using the Poisson brackets
\be  \left\{ K_{aIJ} , \pi^{bKL} \right\} = \delta_a^b \left(\delta_I^K \delta_J^L - \delta_I^L \delta_J^K \right) \text{,} ~~\left\{ \pi^{aIJ}, \pi^{bKL}  \right\} =  \left\{ K_{aIJ}, K_{bKL} \right\} = 0 \ee 
for the extended variables, it has been verified in \cite{BTTI} that ADM Poisson brackets
\be  \left\{ q_{ab}, q_{cd} \right\}_{(K,\pi)} \approx 0 \approx \left\{ P^{ab}, P^{cd} \right\}_{(K,\pi)} \text{,} ~~ \left\{ q_{ab}, P^{cd} \right\}_{(K,\pi)} \approx \delta^{c}_{(a} \delta ^{d}_{b)} \text{,}  \ee 
are reproduced in the extended system up to terms which vanish if the constraints (\ref{eq:gauss},\ref{eq:simplicity}) hold. Actually, only the rotational parts of the Gau{\ss} constraint (\ref{eq:gauss}) is needed for the above Poisson brackets to hold. Without going into the details of the calculation (cf. \cite{BTTI} for further details), we want to point out that (\ref{eq:gauss}) is only needed for the $\{ P, P\}$ - bracket, where terms $K^{[a}\m_{IJ} \pi^{b]IJ} \approx K_{a~K}^{[I} \pi^{aK|J]} \pi^{[b}\m_{IM} \pi^{c]~M}_{~J} \propto \bar{G}^{IJ}$ appear (the bar denotes rotational components, see below for notation). This will become important when proving the validity of the theory with linear simplicity constraint in section \ref{sec:LinearSimplicity}. Since the ADM brackets are recovered, in particular the Dirac algebra of $\mathcal{H}_a$ and $\mathcal{H}$ is reproduced in the extended system, the whole system of constraints $\left\{\mathcal{H}_a,\mathcal{H}, G^{IJ}, S^{ab}_{\overline{M}}\right\}$ can easily be shown to be of the first class \cite{BTTI}.

For later comparison with the solution of the linear simplicity constraints in section \ref{sec:Solution}, we review the solution of the quadratic simplicity constraints as given in \cite{PeldanActionsForGravity, BTTI}. The solution to the quadratic simplicity constraint is in any dimension given by \cite{FreidelBFDescriptionOf, BTTI}
\be
\pi^{aIJ} = 2 n^{[I} E^{a|J]} \text{,}
\label{eq:solsimp}
\ee
where $n^I$ is the unit normal to the vielbein, defined (up to sign) by the equations $n^I n_I = \zeta$ and $n_I E^{aI} = 0$. We can use $n^I$ and the projector $\bar{\eta}_{IJ} := \eta_{IJ} - \zeta n_I n_J $ to decompose any bivector $X_{IJ}$ into its rotational ($\bar{X}_{IJ} := \bar{\eta}_I^K  \bar{\eta}_J^L X_{KL}$) and boost parts (${\bar{X}_I := - \zeta n^J X_{IJ}}$) with respect to $n^I$. Using (\ref{eq:solsimp}), the symplectic potential  reduces to \cite{PeldanActionsForGravity}
\ba
&\m& \frac{1}{2} \pi^{aIJ} \dot{K}_{aIJ}  \nonumber \\ 
&\approx& - \zeta \bar{K}_{aJ} \dot{E}^{aJ} - \bar{K}_{aIJ} E^{aJ} \dot{n}^{I}  \nonumber \\
&\approx& \left[- \zeta \bar{K}_{aJ} - n_J E_{aI} \bar{K}_{bK}\m^{I} E^{bK} \right] \dot{E}^{aJ} \nonumber \\ 
&:=& K'_{aJ} \dot{E}^{aJ} \text{,} \label{eq:SymplecticReduction1}
\ea
where we have dropped total time derivatives and divergences. The inverse vielbein $E_{aI}$ is defined by the equations $E_{aI} E^{a}\m_{J} = \bar{\eta}_{IJ}$ and $E_{aI} E^{bI} = \delta^b_a$ . For the constraints, we find in terms of these variables
\ba
\frac{1}{2} \lambda_{IJ} G^{IJ} &=& - \lambda_{IJ} E^{a[I} K'_a\m^{J]} \text{,} \label{eq:GaussConstraint1} \\
N^a \mathcal{H}_a &\approx& 2 \zeta N^a \nabla_{[a} E^{bI} K'_{b]I} \text{,}  \label{eq:DiffeoConstraint1} \\
N\mathcal{H} &\approx& N\left( \frac{s}{2} E^{aI} E^{bJ} R_{abIJ} - E^{a[I} E^{b|J]} K'_{aI} K'_{bJ} \right) \text{,} \label{eq:HamConstraint1} 
\ea
where terms proportional to the Gau{\ss} constraint (\ref{eq:GaussConstraint1}) as well as total derivatives were dropped in (\ref{eq:DiffeoConstraint1}). Thus we arrive at an already well-established Hamiltonian formulation of General Relativity \cite{PeldanActionsForGravity}, which leads to the ADM formulation after solving the SO$(\eta)$ Gau{\ss} constraint.

\subsubsection{Step 2: $\left\{ \m^{(\beta)}A_{aIJ}, \m^{(\beta)}\pi^{bKL}\right\}$ - Theory}
\label{sec:APi}
Having this extension of the ADM phase space at our disposal, we can turn it into a connection formulation in a second step. We define the spin connection constructed of the $\pi^{aIJ}$ by
\be \Gamma_{aIJ} := \frac{2}{D-1} \pi_{bKL} n^K n_{[I} \partial_a \pi^{b} \m_{J]} \m^L + \zeta \bar{\eta}_{[I}^M \bar{\eta}_{J]K} \pi_{bLM} \partial_a \pi^{bLK} + \zeta \Gamma_{ab}^c \pi^{b} \m_{K[I} \pi_{c|J]} \m^K \text{,} \label{eq:Spinconnection}
\ee
where we used the abbreviations
\ba
  n^I n_J &:=& \frac{1}{D-1} \left( \pi^{aKI} \pi_{aKJ}-\zeta \eta^{I} \m_J \right) \text{,}~~~ \bar{\eta}_{IJ} := \eta_{IJ} - \zeta n_I n_J, ~~\text{and} \nonumber \\
   \pi_{aIJ} &:=& \left(\frac{\zeta}{2} \pi^{aKL} \pi^{b}\m_{KL}\right)^{(-1)} \pi^{b}\m_{IJ} = \frac{1}{q} q_{ab} \pi^{b}\m_{IJ} \text{.} 
  \ea
One can check that $\Gamma_{aIJ}$ satisfies weakly\footnote{Note that when solving the simplicity constraint ($\pi^{aIJ} = 2 n^{[I}E^{a|J]}$), $\Gamma_{aIJ}$ reduces to the hybrid spin connection introduced by Peldan \cite{PeldanActionsForGravity} which annihilates $E^{aI}$ (and, since $n^I$ is a function of $E^{aI}$, also annihilates $n^I$). When additionally choosing time gauge $n^I = \delta^I_0$, the hybrid connection furthermore reduces to the familiar SO$(D)$ spin connection annihilating the SO$(D)$ vielbein $E^{ai}$ ($i = 1,...,D$).} the following identity \cite{BTTI}
\be  \partial_a \pi^{aIJ} + \left[\Gamma_a, \pi^a \right]^{IJ} \approx 0  \text{.}\label{eq:Spinconnection2} \ee 
Moreover, it transforms as a connection under gauge transformations
\be \left\{ \frac{1}{2} G^{KL}\left[\Lambda_{KL}\right], \Gamma_{aIJ} \right\} \approx \partial_a \Lambda_{IJ} +\left[\Gamma_{a},\Lambda\right]_{IJ}  \ee
if the simplicity constraint holds. This suggests the introduction of the following connection variables
\be  \m^{(\beta)} A_{aIJ} := \Gamma_{aIJ} + \beta K_{aIJ} ~~\text{and}~~ \m^{(\beta)}\pi^{aIJ} :=  \frac{1}{\beta} \pi^{aIJ}  \text{,} \ee
with a free parameter\footnote{Since $\Gamma_{aIJ}$ is a homogeneous function of degree zero in $\pi^{aIJ}$ and its derivatives, it is unaffected by the constant rescaling $\pi^{aIJ} \rightarrow \m^{(\beta)}\pi^{aIJ}$.}
 $\beta \in \mathbb{R}/\{0\}$ and Poisson brackets given by
\be  \left\{ \m^{(\beta)}A_{aIJ} , \m^{(\beta)}\pi_{bKL} \right\} = \delta_a^b \left(\delta_I^K \delta_J^L - \delta_I^L \delta_J^K \right) \text{,} ~~\left\{ \m^{(\beta)}\pi^{aIJ}, \m^{(\beta)}\pi^{bKL}  \right\} =  \left\{ \m^{(\beta)}A_{aIJ}, \m^{(\beta)} A_{bKL} \right\} = 0  \text{.} \label{eq:PoissonAPi} \ee
Using equation (\ref{eq:Spinconnection2}), we can rewrite the Gau{\ss} constraint to obtain a familiar expression for the generator of gauge transformations
\be  G^{IJ} = 0 + \left[\m^{(\beta)} K_a, \m^{(\beta)} \pi^a \right]^{IJ} \approx \partial_a \m^{(\beta)} \pi^{aIJ} + \left[\m^{(\beta)}A_a, \m^{(\beta)} \pi^a \right]^{IJ} \text{.} \ee 
Now we can repeat the above analysis, again expressing the ADM variables and constraints in terms of the new ones, 
\be
\label{eq:ExtendedAdmVariables2}
\m^{(\beta)} \pi^{aIJ}\m^{(\beta)} \pi^{b}\m_{IJ} := \frac{2\zeta}{\beta^2} q q^{ab} ~~~\text{and}~~~
\sqrt{q} K_a \m^b := -\frac{s}{4} \m^{(\beta)} \pi^{bIJ} \left( \m^{(\beta)} A-\Gamma \right)_{aIJ} \text{,}
\ee
\ba
 \mathcal{H}_a &=& -\frac{1}{2} \nabla_b \left( \left( \m^{(\beta)} A-\Gamma \right)_{aIJ} \m^{(\beta)}\pi^{bIJ} - \delta_a^b \left( \m^{(\beta)} A-\Gamma \right)_{cIJ} \m^{(\beta)}\pi^{cIJ} \right) \text{,} \\
 \mathcal{H} &=& - \frac{s}{8\sqrt{q}} \left( \m^{(\beta)}\pi^{[a|IJ} \m^{(\beta)}\pi^{b]KL} \left( \m^{(\beta)} A-\Gamma \right)_{bIJ} \left( \m^{(\beta)} A-\Gamma \right)_{aKL}\right) - \sqrt{q} R \label{eq:HamConstraint} \text{,}
\ea
and checking that the ADM Poisson brackets are reproduced on the extended phase space $\{\m^{(\beta)}A,\m^{(\beta)}\pi\}$ up to Gau{\ss} and (quadratic) simplicity constraints \cite{BTTI}. 

From the classical point of view, this formulation is a genuine connection formulation of General Relativity. In the quantum theory, the quadratic simplicity constraint leads to anomalies both in the covariant \cite{EngleLoopQuantumGravity} as well as in the canonical approach \cite{BTTII}. Therefore, we want to introduce a linear simplicity constraint in the canonical theory in the next section, inspired by the new Spin Foam models \cite{EngleTheLoopQuantum, LivineNewSpinfoamVertex, EngleFlippedSpinfoamVertex, EngleLoopQuantumGravity, FreidelANewSpin, KaminskiSpinFoamsFor}.

\subsection{Introducing Linear Simplicity Constraints}
\label{sec:LinearSimplicity}

Recall that the solution to the (quadratic) simplicity constraint in dimensions $D \geq 3$ is given by \cite{FreidelBFDescriptionOf}\footnote{In $D = 3$, an additional topological sector exists \cite{FreidelBFDescriptionOf}. The above results hold in $D = 3$ only if this sector is excluded by hand.} $S^{ab}_{\overline M} = 0 \Leftrightarrow \m^{(\beta)}\pi^{aIJ} = \frac{2}{\beta} n^{[I} E^{a|J]} $ and that $n^I$ is no independent field but determined by the vielbein $E^{aI}$. We now postulate a new field $N^I$, which will play the role of this normal, together with its conjugate momentum $P_I$, subject to the linear simplicity and normalisation constraints
\ba
S^{a}_{I \overline M} &:=& \epsilon_{IJKL\overline M} ~ N^J ~ \m^{(\beta)}\pi^{a KL} \text{,} \label{eq:LinearSimplicity} \\
\mathcal{N} &:=& N^I N_I - \zeta \text{.} \label{eq:Normalization}
\ea
The solution to the linear simplicity constraints in any dimension $D \geq 3$ is given by\footnote{Using the linear simplicity constraints, we automatically exclude the topological sector in $D = 3$.} $\m^{(\beta)}\pi^{aIJ} = \frac{2}{\beta} N^{[I} E^{a|J]}$, with $N_I E^{aI} = 0$. We see that on the solutions, the physical information of $\m^{(\beta)}\pi^{aIJ}$ is encoded in the vielbein $E^{aI}$, which in turn fixes the direction of $N^I$ completely. The remaining freedom in choosing its length is fixed by the normalisation constraint $\mathcal{N}$ and we find $N^I = n^I(E)$, i.e. the $N^I$ are no physical degrees of freedom. The same has to be assured for the momenta $P^I$, i.e. we should add additional constraints $P^I = 0$. However, these extra conditions can be interpreted as (partial) gauge fixing conditions for (\ref{eq:LinearSimplicity},\ref{eq:Normalization}), which then can be removed by applying the procedure of gauge unfixing. We will take a short-cut and directly ``guess'' the theory such that the constraints (\ref{eq:LinearSimplicity},\ref{eq:Normalization}) are implemented as first class, and we will show that when solving these constraints, the momenta $P_I$ are automatically removed from the theory.
	
The theory we want to construct is very similar to the $\{\m^{(\beta)} A_{aIJ} , \m^{(\beta)} \pi^{bKL} \}$ - theory of section \ref{sec:APi}. It is defined by the Poisson brackets (\ref{eq:PoissonAPi}) and 
\be  \left\{ N^I, P_J \right\} = \delta_J^I \text{,} ~~\left\{ N^I,N^J \right\} =  \left\{ P_I,P_J \right\} = 0  \text{,} \ee
and, apart from the linear simplicity and normalisation constraints (\ref{eq:LinearSimplicity},\ref{eq:Normalization}), is subject to
\ba
G^{IJ} &=& \frac{1}{2} \m^{(\beta)}D_a \m^{(\beta)}\pi^{aIJ} + P^{[I}N^{J]} \text{,} \\
\mathcal{H}_a &=& \frac{1}{2} \m^{(\beta)}\pi^{bIJ} \partial_a \m^{(\beta)}A_{bIJ} - \frac{1}{2} \partial_b \left( \m^{(\beta)}\pi^{bIJ} \m^{(\beta)}A_{aIJ} \right) + P_I \partial_a N^I \text{,} \label{eq:VectorConstraint} \\
\mathcal{H} &=& - \frac{s}{8\sqrt{q}} \left[ \m^{(\beta)}\pi^{[a|IJ} \m^{(\beta)}\pi^{b]KL} \left(\m^{(\beta)}A-\Gamma\right)_{bIJ} \left(\m^{(\beta)}A-\Gamma\right)_{aKL}\right] - \sqrt{q} R \text{.}
\ea
Note that the Hamilton constraint is the same\footnote{In particular, we want to point out that it is not the Hamilton constraint for gravity coupled to standard scalar fields $\phi$, which would obtain additional terms $\sim \frac{p^2}{\sqrt{\det{q}}} + \sqrt{\det{q}} q^{ab} \phi_{,a} \phi_{,b}$ for the scalar field $\phi$ and its conjugate momentum $p$ which are missing here. In fact, these terms would spoil the constraint algebra, since $\{\mathcal{H}, S^a_{I\overline{M}}\}$ and $\{\mathcal{H}, \mathcal{N}\}$ would not vanish weakly.} as in equation (\ref{eq:HamConstraint}), whereas the Gau{\ss} and vector constraint differ and are chosen such that they obviously generate SO$(\eta)$ gauge transformations and spatial diffeomorphisms respectively on all phase space variables. In the following, we prove its equivalence to the ADM formulation. First of all, we will show that the Poisson brackets of the ADM variables $K^{ab}$, $q_{ab}$ in terms of the new variables $\m^{(\beta)}A_{aIJ}$, $\m^{(\beta)}\pi^{aIJ}$, $N^I$, and $P_I$ are still reproduced on the extended phase space up to constraints. This is non-trivial, even if the expressions for $K^{ab}(\m^{(\beta)}\pi,\m^{(\beta)}A)$, $q_{ab}(\m^{(\beta)}\pi)$ are given by (\ref{eq:ExtendedAdmVariables2}) as in the previous section, since we changed both the simplicity and the Gau{\ss} constraint. For the linear simplicity and normalisation constraints, the solution for $\m^{(\beta)}\pi^{aIJ}$ is the same as in the case of the quadratic simplicity (neglecting the topological sector), $\m^{(\beta)}\pi^{aIJ} = \frac{2}{\beta} n^{[I}E^{a|J]}$, and terms which vanished due to the quadratic simplicity constraint still vanish in the case at hand. For the Gau{\ss} constraint, note that the only terms appearing in the calculation are of the form $(\m^{(\beta)}A-\Gamma)^{[a}\m_{IJ}\m^{(\beta)}\pi^{b]IJ}$, which already vanish on the surface defined by the vanishing of the rotational parts of the Gau{\ss} constraint (cf. section \ref{sec:KPi}). Now, if the linear simplicity and normalisation constraints hold, we know that $N^I = n^I(E)$, i.e. the modification of the Gau{\ss} constraint $P^{[I}N^{J]} \approx \bar{P}^{[I}n^{J]}$ on-shell just changes the boost part of the Gau{\ss} constraint. Thus, the ADM brackets are reproduced on the surface defined by the vanishing of $G^{IJ}$, $S^{a}_{I \overline M}$ and $\mathcal{N}$.

Next, we will show that the algebra is of  first class. Note that since $G^{IJ}$ and $\mathcal{H}_a$ generate gauge transformations and spatial diffeomorphisms by inspection, their algebra with all other constraints obviously closes. The algebra of the linear simplicity and the normalisation constraint is trivial. Moreover, the Hamilton constraint Poisson-commutes trivially with the normalisation constraint and, since it depends only on the combination $\m^{(\beta)}\pi^{aIJ} \m^{(\beta)}A_{bIJ}$, we find $\left\{\mathcal{H}[N], S^{a}_{I\overline{M}}[s_a^{I\overline{M}}]\right\} = S^{a}_{I\overline{M}}[...]$. We are left with the Poisson-bracket between two Hamilton constraints. Since on-shell the ADM brackets are reproduced, the result is
\be
\left\{ \mathcal{H}[M], \mathcal{H}[N]\right\} \approx \mathcal{H}'_a[q^{ab}\left(MN_{,b} - NM_{,b}\right)] \text{,}
\ee
where $\mathcal{H}'_a = -2 q_{ac} \nabla_b P^{bc}$ now denotes the ADM diffeomorphism constraint. As we will see in the next section \ref{sec:Solution}, the vector constraint (\ref{eq:VectorConstraint}) correctly reduces to the ADM diffeomorphism constraint if the Gau{\ss} and simplicity constraint hold, $\mathcal{H}_a \approx \mathcal{H}_a'$. Therefore, the algebra closes. What is left to show is that also the Hamilton constraint $\mathcal{H}$ on-shell reproduces the ADM constraint, which will be made explicit when solving the constraints in the next section \ref{sec:Solution}. Note that because of the modified Gau{\ss} and simplicity constraint, again this is non-trivial even if $\mathcal{H}$ is identical with (\ref{eq:HamConstraint}).

The counting of the number of physical degrees of freedom goes as follows: The full phase space consists of $\left|\left\{A,\pi,N,P\right\}\right| = 2 \frac{D^2(D+1)}{2} + 2 (D+1)$ degrees of freedom which are subject to \mbox{$\left|\left\{\mathcal{H}_a,\mathcal{H},G^{IJ},S^{a}_{I \overline M},\mathcal{N}\right\}\right|$} = $(D+1) + \frac{D(D+1)}{2} + \frac{D^2(D-1)}{2} + 1$ first class constraints. It is most convenient to compare this to Peldan's \cite{PeldanActionsForGravity} extended ADM formulation given at the end of section \ref{sec:KPi}, with $\left|\left\{E,K\right\}\right| = 2D(D+1)$ phase space degrees of freedom and the first class constraints $\left|\left\{\mathcal{H}_a,\mathcal{H},G^{IJ}\right\}\right| = (D+1) + \frac{D(D+1)}{2}$. In any dimension, the difference in phase space degrees of freedom is exactly removed by the simplicity and normalisation constraint, $\left|\left\{A,\pi,N,P\right\}\right| - \left|\left\{E,K\right\}\right| = 2 \left|\left\{S^{a}_{I \overline M},\mathcal{N}\right\}\right|$.

We remark that related formulations of General Relativity, where a time normal appears as an independent dynamical field, have already appeared in the literature \cite{SaHamiltonianAnalysisOf, AlexandrovSU(2)LoopQuantum, GeillerALorentzCovariant}. The difference between these and our formulation is that while our formulation features both the simplicity constraint and the time normal at the same time, the time normal appears in the process of solving the simplicity constraint without solving the boost part of the Gau{\ss} constraint in the other approaches. In other words, the time normal is an integral part of the simplicity constraint in our approach, not a concept emerging after its solution.

\subsection{Classical Solution of the Linear Simplicity Constraints}
\label{sec:Solution}
Solving the linear simplicity and normalisation constraints can be done similarly as in section \ref{sec:KPi}. As already pointed out in section \ref{sec:LinearSimplicity}, solving these constraints leads to
\be
\m^{(\beta)}\pi^{aIJ} = \frac{2}{\beta} n^{[I} E^{a|J]} ~~~\text{and}~~~ N^I = n^I(E) \text{.} 
\ee
We make the Ansatz $\m^{(\beta)}A_{aIJ} = \Gamma_{aIJ} + \beta K_{aIJ}$ with $\Gamma_{aIJ}$ defined as in (\ref{eq:Spinconnection}). Then, the symplectic potential reduces to \cite{PeldanActionsForGravity}
\ba
&\m& \frac{1}{2} \m^{(\beta)}\pi^{aIJ} \m^{(\beta)}\dot{A}_{aIJ} + P_I \dot{N}^I \nonumber \\ 
&\approx& \bar{K}_{aJ} \dot{E}^{aJ} - \bar{K}_{aIJ} E^{aJ} \dot{n}^{I} + \bar{P}_I \dot{n}^I \nonumber \\
&\approx& \left[ \bar{K}_{aJ} - n_J E_{aI} \left( \bar{K}_{bK}\m^{I} E^{bK} + \bar{P}^I\right)\right] \dot{E}^{aJ} \nonumber \\ 
&:=& K''_{aJ} \dot{E}^{aJ} \text{,}
\ea
where we have dropped total time derivatives and divergences. Note that, compared to (\ref{eq:SymplecticReduction1}), the result is the same up to the additional $\bar{P}^I$ term appearing in the definition of $K''_{aI}$. For the constraints, we find equal expressions as in (\ref{eq:GaussConstraint1}, \ref{eq:DiffeoConstraint1}, \ref{eq:HamConstraint1}) with $K'_{aI}$ replaced by $K''_{aI}$ and again arrive at Peldan's extended ADM formulation without time gauge \cite{PeldanActionsForGravity}, which leads to the ADM formulation after solving the SO$(\eta)$ Gau{\ss} constraint.

\subsection{Linear Simplicity Constraints for Theories with Dirac Fermions}
\label{sec:Fermions}
In order to incorporate fermions into the framework, tetrads and their higher dimensional analogues (vielbeins) have to be used at the Lagrangian level to construct a representation of the spacetime Clifford algebra. Therefore, the extension of the ADM phase space introduced above is not applicable here. In \cite{BTTIV} it is shown that the symplectic reduction of the extension of the phase space $\left(K_{ai}, E^{bj}\right)$ with SO$(D)$ Gau{\ss} constraint to $\left(A_{aIJ},\pi^{bKL}\right)$ with SO$(\eta)$ Gau{\ss} and (quadratic) simplicity constraint gives back the 
unextended theory. We may now apply this fact to theories with fermions. The explicit construction is 
\be
\bar{E}^{aI} = \zeta \bar{\eta}^I\m_J\pi^{aKJ}n_K\text{,}~~~~ \bar{K}_{aI} = \zeta \bar{\eta}_I\m^J(A-\Gamma)_{aKJ}n^K \text{,}
\ee
where $\Gamma_{aIJ}$, $\bar\eta_{IJ}$ and $n^I$ are understood as functions of $\pi^{aIJ}$ (cf. \cite{BTTIV} for more details).

The extension of $\left(K_{ai}, E^{bj}\right)$ with SO$(D)$ Gau{\ss} constraint to $\left(A_{aIJ},\pi^{bKL}, N^I, P_J \right)$ with SO$(\eta)$ Gau{\ss}, linear simplicity and normalisation constraint works exactly the same way. We can even choose to simplify the replacement of the vielbein and extrinsic curvature using the normal $N^I$,
\be
\bar{E}^{aI} = \zeta \bar{\eta}^I\m_J\pi^{aKJ}N_K\text{,}~~~~ \bar{K}_{aI} = \zeta \bar{\eta}_I\m^J(A-\Gamma)_{aKJ}N^K \text{,}	
\ee
where $\bar\eta_{IJ}$ now is understood as a function of $N^I$. The calculations are completely analogous to those in \cite{BTTIV} and therefore will not be detailed here.

\subsection{Kinematical Hilbert Space for $N^I$}
\label{sec:KinematicalHilbertSpace}
We restrict to the case $\zeta = 1$ in the following, because the kinematical Hilbert space for canonical Loop Quantum Gravity has been defined rigorously only for compact gauge groups like SO$(D+1)$. For scalar fields like the Higgs field, two different constructions to obtain a kinematical Hilbert space have been given. In the first one \cite{ThiemannKinematicalHilbertSpaces}, a crucial role is played by point holonomies $U_x(\Phi) := \exp\left(\Phi^{IJ}(x) \tau_{IJ}\right)$. The field $\Phi^{IJ}$, which is assumed to transform according to the adjoint representation of $G$, is contracted with the basis elements $\tau_{IJ}$ of the Lie algebra of $G$ and then exponentiated. Point holonomies are better suited for background independent quantisation than the field variables $\Phi^{IJ}$ themselves, since the latter are real valued rather than valued in a compact set. Therefore, a Gau{\ss}ian measure would be a natural choice for the inner product for $\Phi^{IJ}$, but this is in conflict with diffeomorphism invariance \cite{ThiemannKinematicalHilbertSpaces}. In the case at hand, this framework is not applicable, since $N^I$ transforms in the defining representation of SO$(D+1)$ and therefore, there is no (or, at least no obvious) way to construct point holonomies from $N^I$ in such a way that the exponentiated objects transform ``nicely" under gauge transformations. The second avenue \cite{ThiemannModernCanonicalQuantum} for background independent quantisation of scalar fields leads to a diffeomorphism invariant Fock representation and can be applied in principle. However, in the case at hand there is a more natural route. On the constraint surface $\mathcal{N} = N^I N_I - 1 = 0$, $N^I$ is valued in the compact set $S^{D}$. In this case the measure problems can be circumvented by solving $\mathcal{N}$ classically. The
obvious  choice of Hilbert space is then of course the space of square integrable functions on the $D$-sphere.\\
\\
 To solve $\mathcal{N}$, we choose a second class partner $\tilde{\mathcal{N}} := N^I P_I$, 
\be
\left\{N^I(x) N_I(x) - 1,  N^J(y) P_J(y) \right\} = N^I(x) N_I(x) \delta^D(x-y) \approx \delta^D(x-y)\text{,}
\ee 
where terms $\propto \mathcal{N}$ have been dropped. $\tilde{\mathcal{N}}$ weakly Poisson commutes with the constraints: it is Gau{\ss} invariant and transforms diffeomorphism covariant, it trivially Poisson commutes with $\mathcal{H}$ (which neither depends on $N^I$ nor on $P_I$), and a short calculations yields
\be
\left\{S^a_{I \overline{M}}[s_a^{I\overline{M}}], \tilde{\mathcal{N}}[\tilde{n}]\right\} =  S^{a}_{I\overline{M}}[\tilde{n} s_a^{I\overline{M}}] \text{.}
\ee
Therefore, the algebra of the remaining constraints is unchanged when we solve $\mathcal{N}$, $\tilde{\mathcal{N}}$ using the Dirac bracket. Let $\bar{\eta}_{IJ}=\eta_{IJ}-N_I N_J/||N||^2$ whence 
$\bar{\eta}_{IJ} N^J=0$ also when $||N||\not=1$. Then $\bar{P}_I=\bar{\eta}_{IJ} P^J$ 
Poisson commutes 
with the normalisation constraint and thus is an observable just as $N^I$. Since the
Dirac matrix is diagonal, the 
Dirac brackets of $\bar{P}_I,N^I$ coincide with their Poisson brackets. We find  
\begin{eqnarray}
\left\{N^I(x), N_J(y)\right\}_{DB} &=& 0\text{,} \nonumber \\ 
 \left\{N^I(x), \bar{P}_J(y)\right\}_{DB} &=& \bar{\eta}^I\m_J(x) \delta^{D}(x-y) \text{,} \nonumber \\
 \left\{\bar{P}^I(x), \bar{P}^J(y)\right\}_{DB} &=& 2 \bar{P}^{[I}(x)N^{J]}(x) \delta^{D}(x-y)\text{,}
\end{eqnarray}
while the remaining brackets are unaffected. We see that unfortunately the Poisson algebra 
of the $N^I$ and $\bar{P}_I$ does not close, it automatically generates also the rotation generator 
$L_{IJ}=2 N_{[I }P_{J]}=2 N_{[I} \bar{P}_{J]}$. We therefore have to include it into our 
algebra. On the other hand obviously $\{L_{IJ},\mathcal{N}\}=0$ so that $L_{IJ}$ is also an observable
and moreover the $L_{IJ}$ generate the Lie algebra so$(D+1)$ while $\{L_{IJ},N_K\}
=-2 N_{[I} \delta_{J]K}$ so that the algebra of the $N_I,L_{IJ}$ already closes. Finally we have the 
identity $L_{IJ} N^J=-||N||^2 \bar{P}_I$ so that the $N^I, L_{IJ}$ already determine 
$\bar{P}_I$. We conclude that nothing is gained by considering the $\bar{P}_I$ and that 
it is better to consider the overcomplete set of observables $N^I, L_{IJ}$ instead.\\
\\
Consider, similar as in LQG, cylindrical functions $F[N]$ of the form 
$F[N]=F_{p_1,..,p_n}(N(p_1),..,N(p_n))$ where $F_{p_1.. p_n}$ is a 
polynomial with complex coefficients of the $N^I(p_k),\;k=1,..,n;\;I=0,..,D+1$. 
We define the operator $\hat{N}_I(x)$ to be multiplication by $N_I(x)$ on this space.
Let 
also $\Lambda^{IJ}$ be a smooth antisymmetric matrix valued function of 
compact support and define the operator 
\be \label{a}
\hat{L}[\Lambda]:=2\int\; d^Dx \; \Lambda^{IJ}(x) \; \hat{N}_{[I}(x)\; \hat{P}_{J]}(x) \text{,}
\ee
where $\hat{P}_J(x)=i\delta/\delta N_J(x)$. Notice that no factor ordering problems arise.
The operator $\hat{L}[\Lambda]$ has a well defined action on cylindrical functions, specifically
\be \label{b}
\hat{L}[\Lambda] \; F[N]=2i\sum_{k=1}^n\; 
\Lambda^{IJ}(p_k) \hat{N}_{[I}(p_k) \partial/\partial N_{J]}(p_k)\; F[N]
\ee     
and annihilates constant functions. 

Let $d\nu(N):=c_D d^{D+1}N \delta(||N||^2-1)$ the SO$(D+1)$ invariant measure on $S^D$ where 
the constant $c_D$ makes it a probability measure. For a function cylindrical over
the finite point set $\{p_1,..,p_n\}$ we define the following positive linear functional
\be \label{bb}
\mu[F]:=\int\; d\nu(N_1)\;..\;d\nu(N_n)\; F_{p_1..p_n}(N_1,..,N_n) \text{.}
\ee
Just as in LQG it is straightforward to show that the measure is consistently defined and thus 
has a unique $\sigma-$additive extension to the projective limit of the finite Cartesian products 
of copies of $S^D$ which in this case is just the infinite Cartesian product
 $\overline{{\cal N}}:=\prod_x S^D$
of copies of $S^D$ \cite{YamasakiMeasuresOnInfinite}, one for each spatial point. This space can be considered as 
a space of distributional normals because a generic point in it is a collection
of vectors $(N(x))_x$ without any continuity properties. The operator $\hat{N}_I(x)$ is bounded and trivially self-adjoint because
$N_I(x)$ is real valued and $S^D$ is compact. To see that $\hat{L}[\Lambda]$ is self adjoint
we let $g_\Lambda(p)=\exp(\Lambda^{IJ}(p)\tau_{IJ})$ where $\tau_{IJ}$ are the generators of 
so$(D+1)$. We define the operator 
\be \label{c}
\left( \hat{U}(\Lambda) F \right)[N]:=F_{p_1..p_n} \left(g_\Lambda(p_1) N(p_1),..,g_\Lambda(p_n) N(p_n)\right)  \text{,}
\ee
which can be verified to be unitary and strongly continuous in $\Lambda$. It maybe verified 
explicitly that 
\be \label{d}
\hat{L}[\Lambda]=\frac{1}{i}\; [\frac{d}{dt}]_{t=0} \hat{U}[t\Lambda] \text{,}
\ee
whence $\hat{L}[\Lambda]$ is self-adjoint by Stone's theorem \cite{ReedBook1}. 
Finally it is straightforward to check that besides the $^\ast$-relations also the commutator 
relations hold, i.e. they reproduce $i$ times the classical Poisson bracket.\\
\\
We conclude that we have found a suitable background independent representation
of the normal field sector. \\
\\
At each point $p \in \Sigma$, an orthonormal basis in the Hilbert space 
$\mathcal{H}_p=L_2(S^D,d \nu)$ is given by the generalisations of spherical harmonics to higher dimensions $\Xi_l^{\vec K}(N) $, which are shortly introduced in the appendix of our companion paper \cite{BTTV} (see \cite{VilenkinSpecialFunctionsAnd} for a comprehensive treatment). An orthonormal basis for $\mathcal{H}_N$ is given by spherical harmonic vertex functions $F_{\vec{v}, \vec{l}, \vec{\vec{K}}}(N) : = \prod_{v \in \vec v} \Xi_{l_v}^{\vec{K}_v}(N)$. Any cylindrical function $F_{\vec v}$ can be written as a mean-convergent series of the form
\be
	F_{\vec v}(N) = \sum_{\vec{l},\vec{\vec K}} a_{\vec{v}, \vec{l}, \vec{\vec K}} F_{\vec{v}, \vec{l}, \vec{\vec{K}}}(N)
\ee
for complex coefficients $a_{\vec{v},\vec{l},\vec{\vec{K}}}$. The sum here runs for each $v \in \vec{v}$ over all values $l \in \mathbb{N}_0$ and for each $l$ over all $\vec{K}$ compatible with $l$. Following the construction in \cite{ThiemannKinematicalHilbertSpaces} we obtain the combined Hilbert space for the scalar field and the connection simply by the tensor product, $\mathcal{H}_T = \mathcal{H}_{\text{grav}} \otimes \mathcal{H}_N = L_2(\overline{\mathcal{A}}^{\text{SO}(D+1)}, d\mu_{AL}^{\text{SO}(D+1)}) \otimes L_2(\overline{{\cal N}}, d\mu_N)$. An orthonormal basis in this space is given by a slight generalisation of the usual gauge-variant spin network states (cf., e.g., \cite{ThiemannKinematicalHilbertSpaces}), where each vertex is labelled by an additional simple SO$(D+1)$ irreducible representation coming from the normal field. This of course leads to an obvious modification of the definition of the intertwiners which also have to contract the indices coming from this additional representation.

\end{appendix}

\newpage

\bibliography{pa92pub.bbl}

\end{document}